\documentclass[preprint,superscriptaddress,nofootinbib,showkeys,showpacs,tightenlines]{revtex4}  
\usepackage{graphicx}
\usepackage{dcolumn}
\newcommand{\be}{\begin{equation}}
\newcommand{\ee}{\end{equation}}
\newcommand{\bee}{\begin{eqnarray}}
\newcommand{\eee}{\end{eqnarray}}
\newcommand{\eq}{\end{quote}}
\newcommand{\nn}{\nonumber}
\newcommand{\Slash}[1]{\ooalign{\hfil/\hfil\crcr$#1$}}

\def\gsim{\displaystyle\mathop{>}_{\sim}}
\def\lsim{\displaystyle\mathop{<}_{\sim}}
\preprint{PNU-NTG-10/2005}
\begin{document}      
\title{Production of $\Lambda(1520)$}
\author{Seung-Il Nam}
\email{sinam@rcnp.osaka-u.ac.jp}
\affiliation{Research Center for Nuclear Physics (RCNP), Osaka
  University, Ibaraki, Osaka
567-0047, Japan}
\affiliation{Department of Physics and Nuclear Physics \& Radiation
Technology Institute (NuRI), 
Pusan National University, Keum-Jung~gu, Busan 609-735, Republic of Korea} 
\author{Atsushi Hosaka}
\email{hosaka@rcnp.osaka-u.ac.jp}
\affiliation{Research Center for Nuclear Physics (RCNP), Osaka
  University, Ibaraki, Osaka
567-0047, Japan}
\author{Hyun-Chul Kim}
\email{hchkim@pusan.ac.kr}
\affiliation{Department of Physics and Nuclear Physics \& Radiation Technology Institute (NuRI),
Pusan National University, Keum-Jung~gu, Busan 609-735, Republic of Korea} 

\begin{abstract}
We investigate the $\Lambda(1520)$ photoproduction via the reaction
process, $\gamma N\to K\Lambda(1520)$. We employ the Born
approximation and the Rarita-Schwinger formalism is used for
$\Lambda(1520)$. We reproduce the total cross sections and the various 
angular distributions qualitatively well for the proton target and
estimate them for the 
neutron one. We find that the contact term contributes much more to
the process than the other kinematical channels. Taking into account
this fact, we reanalyze the  
$K^*$--exchange dominance hypothesis suggested by the previous experiments. 
\end{abstract}
\maketitle
\section{Introduction}
The observation of the evidence of the exotic pentaquark baryon
$\Theta^+$ has been one of the hottest issues in recent years in hadron
physics~\cite{experiment}. However, there have been also many
negative opinions and results for that evidence. Recently, the LEPS
collaboration reported a new positive result for the evidence of
the $\Theta^+$ baryon from the deuteron
target~\cite{Nakano:chiral05}. Interestingly, the production of
$\Theta^+$ took place together with $\Lambda(1520)\,(\equiv\Lambda^*)$
production. Therefore, it is 
natural to expect that there is a deep correlation between the
productions of the 
two baryons, and it can be possible to extract information for the
$\Theta^+$ baryon by analyzing $\Lambda^*$ instead. 

At present, there are several experiments related to the
photoproduction (A.~Boyarski {\it et  
al.}~\cite{Boyarski:1970yc} and D.~P.~Barber {\it et
al.}~\cite{Barber:1980zv}) and the electroproduction (S.~P.~Barrow
{\it et al.}~\cite{Barrow:2001ds}) of $\Lambda^*$. It was reported 
that the photoproduction of $\Lambda^*$ from the proton target was dominated by
vector $K^*$--exchange by analyzing the $\Lambda^*$ decay in the $t$--channel
helicity frame~\cite{Barber:1980zv}. On the other hand, it was claimed
that the pseudoscalar
$K$--exchange was the dominant contribution in the
electroproduction~\cite{Barrow:2001ds}. 

In the present report, we investigate the $\Lambda^*$ photoproduction
via the reaction process $\gamma N \to K\Lambda^*$. As a frame work,
we make use of the Born approximation which is expected to work
properly in the low 
energy region. The Rarita-Schwinger vector-spinor formalism is 
introduced to treat spin--3/2 baryon ($\Lambda^*$)
relativistically. A phenomenologically parameterized form factor is taken
into account in gauge invariant manner. 

We obtain qualitatively well reproduced total cross sections and 
momentum transfer $t$--dependence for the
proton target case ($\gamma p\to K^+\Lambda^*$) and estimate them for
the neutron one ($\gamma n\to K^0\Lambda^*$). The contact term
contribution is found to be the most 
dominant contribution among various kinematical channels. This fact leads
to the large difference in the order of magnitudes of the total cross
sections for the proton and neutron targets
($\sigma_{p}\gg\sigma_{n}$). Vector $K^*$--exchange plays an 
important role to produce a characteristic angular distribution for the
neutron target case. The $K^*$--exchange dominance hypothesis in the
$\Lambda^*$ photoproduction from the proton
target~\cite{Barber:1980zv} is reanalyzed by decomposing final spin
state of the reaction. We observe that though spin(helicity)-3/2 final state is
important as concluded by the previous analysis~\cite{Barber:1980zv},
the contribution of the 
spin--3/2 comes not only from the $K^*$--exchange but also from the
contact term. Furthermore, we show the contribution from the contact
term is larger than that from the $K^*$--exchange. 

The present report will be organized as follows. In section 2 we
will present the formalism for the present reaction
calculations. section 3 will provide numerical results
including total and 
differential cross sections and the $t$--dependence for the proton and
neutron targets. A reanalysis of the $K^*$--exchange dominance
hypothesis will be given in section 4. Finally, we will summarize our
results and make a brief conclusion in section 5.      
\section{Formalism}
\begin{figure}[tbh]
\includegraphics[width=12cm]{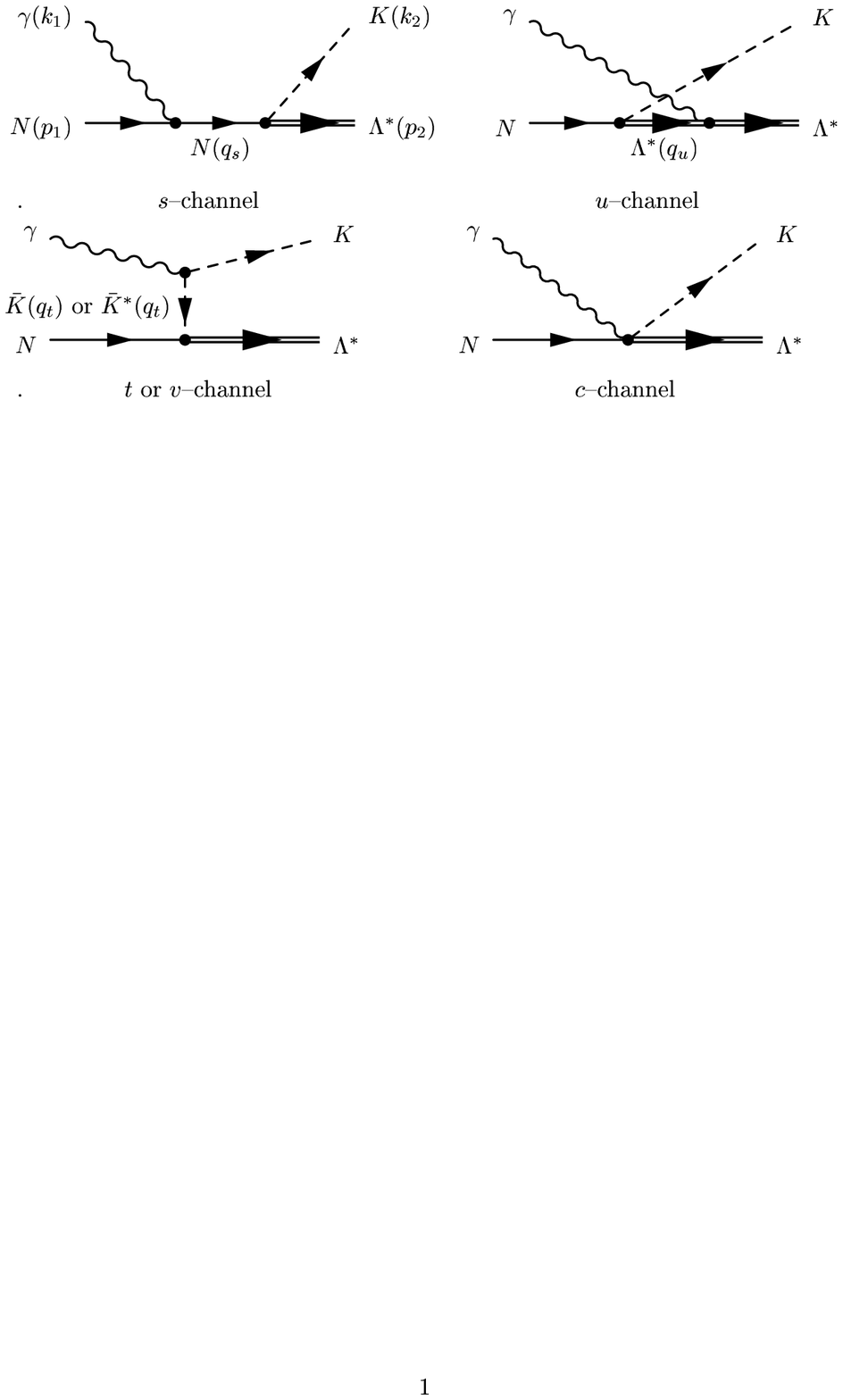}
\caption{The Feynmann diagrams}
\label{fig0}
\end{figure}

We begin with the effective Lagrangians relevant to the $\gamma N\to
K\Lambda^{*}$ process depicted in Fig.~\ref{fig0}. We define the
momenta of photon, pseudo-scalar kaon, vector kaon, nucleon and
$\Lambda^*$ as shown in the figure. For convenience, the vector $K^*$--exchange
in the $t$--channel and contact diagrams will be called as the
$v$--channel (vector channel) and $c$--channel (contact term channel),
respectively. We need to consider all diagrams shown in
Fig.~\ref{fig0} for the proton 
target, whereas only the magnetic (tensor) term 
of the $s$--channel, the $K^*$--exchange ($v$--channel) and the
$u$--channel are necessary for the neutron target due to charge
neutrality. In order to formulate the effective 
Lagrangians including 
spin-3/2 particles, we employ the Rarita-Schwinger (RS) field
formalism which we summarize in the Appendix.  

The relevant effective Lagrangians are given as :  
\bee
\mathcal{L}_{\gamma NN}
&=&-e\bar{p}\left(\gamma_{\mu}+i\frac{\kappa_{p}}{2M_{p}}\sigma_{\mu\nu}k^{\nu}_{1}\right) 
A^{\mu}N\,+{\rm
h.c.},\nn\\
\mathcal{L}_{\gamma KK}&=&
ie\left\{ 
(\partial^{\mu}K^{\dagger})K-(\partial^{\mu}K)K^{\dagger}
\right\}A_{\mu},\nn\\
\mathcal{L}_{\gamma
\Lambda^*\Lambda^*}&=&-\bar{\Lambda^*}^{\mu}
\left\{\left(-F_{1}\Slash{\epsilon}g_{\mu\nu}+F_3\Slash{\epsilon}\frac{k_{1
\mu}k_{1
\nu}}{2M^{2}_{\Lambda^*}}\right)-\frac{\Slash{k}_{1}\Slash{\epsilon}}
{2M_{\Lambda^*}}\left(-F_{2}g_{\mu\nu}+F_4\frac{k_{1\mu}k_{1 \nu}}
{2M^{2}_{\Lambda^*}}\right)\right\}\Lambda^{*\nu}\nn\\&+&{\rm 
h.c.},\nn\\
\mathcal{L}_{\gamma
  KK^{*}}&=&g_{\gamma
  KK^{*}}\epsilon_{\mu\nu\sigma\rho}(\partial^{\mu}A^{\nu})
(\partial^{\sigma}K)K^{*\rho}\,+{\rm 
h.c.},\nn\\
\mathcal{L}_{KN\Lambda^*}&=&\frac{g_{KN\Lambda^*}}{M_{K}}\bar{\Lambda^*}^{\mu}
\Theta_{\mu\nu}(A,Z)(\partial^{\nu}K){\gamma}_{5}p\,+{\rm 
h.c.},\nn\\
\mathcal{L}_{K^{*}N\Lambda^*}&=&-\frac{ig_{K^{*}N\Lambda^*}}
{M_{V}}\bar{\Lambda^*}^{\mu}\gamma^{\nu}(\partial_{\mu}
K^{*}_{\nu}-\partial_{\nu}K^{*}_{\mu})p+{\rm 
h.c.},\nn\\ \mathcal{L}_{\gamma
KN\Lambda^*}&=&-i\frac{eg_{KN\Lambda^*}}{M_{K}}\bar{\Lambda^*}^{\mu}A_{\mu}K{\gamma}_{5}N\,+{\rm
h.c.},
\label{Lagrangian}
\eee
where $N$, $\Lambda^*_{\mu}$, $K$ and $A^{\mu}$ are the nucleon,
$\Lambda^*$, pseudoscalar kaon and photon 
fields, respectively. The interaction for $K^{*}N\Lambda^*$ vertex is taken   
from Ref.~\cite{Machleidt:1987hj}.  As for the $\gamma \Lambda^* \Lambda^*$ vertex
for the $u$--channel, we utilize the effective interaction suggested
by Ref.~\cite{gourdin} which contains four form factors of different
multipoles. We ignore the electric coupling $F_1$, since $\Lambda^*$
is neutral. We also neglect $F_3$ and $F_4$ 
terms, assuming that higher multipole terms are less important. Hence,
for the photon coupling to $\Lambda^*$, we consider only the magnetic
coupling term $F_2$ whose strength is proportional to the anomalous
magnetic moment of $\Lambda^*$, $\kappa_{\Lambda^*}$ which is treated
as a free parameter. The off-shell term $\Theta_{\mu\nu}(A,Z)$ of a
general spin-3/2 particle is defined as follows~\cite{Nath:wp,Hagen:ea} : 
\be
\Theta_{\mu\nu}(A,Z) = g_{\mu\nu}+\left\{\frac{1}{2}
\left(1+4Z)A+Z\right)\right\}\gamma_{\mu}\gamma_{\nu}. 
\label{theta} 
\ee
If we choose $A=-1$~\cite{Read:ye,Nath:wp,Hagen:ea}, we can
rewrite Eq.~(\ref{theta}) in the following form with new parameter
$X=-(Z+1/2)$ :
\be
\Theta_{\mu\nu}(X)=g_{\mu\nu}+X\gamma_{\mu}\gamma_{\nu}.
\ee
Here, we will regard $X$ as a free parameter in the present work. 

In
order to determine the coupling constant $g_{KN\Lambda^*}$, we make
use of the full width $\Gamma_{\Lambda}  = 15.6$ MeV and the branching
ratio 0.45 for the decay $\Lambda^*\to
\bar{K}N$~\cite{Eidelman:2004wy}. The coupling constant $KN\Lambda^*$
can be obtained by the following relation :
\bee
g_{KN\Lambda^*}=\left\{\frac{P_{3}}{4\pi
  M^{2}_{\Lambda^{*}}M^{2}_{K}{\Gamma}_{\Lambda^{*}}}\left(\frac{1}{4}
\sum_{\rm
spin}|\mathcal{M'}|^{2}\right)\right\}^{-\frac{1}{2}},\,\,\,\,\,i\mathcal{M'}  
&=&\bar{u}(P_2)\gamma_{5}P_{3}^{\mu}u_{\mu}(P_1),\nn\\
\label{coupling}
\eee
where $P_1$, $P_2$ and $P_3$ are the momenta of $\Lambda^{*}$, $N$
and $\bar{K}$, respectively  for the two body decay 
$\Lambda^{*}\to\bar{K}N$ in the center of mass frame. Thus, we obtain
$g_{KN\Lambda^*}\sim 11$. As for the $K^*N\Lambda^*$ coupling
constant, we will choose the values of $|g_{K^*N\Lambda^*}|=0$ and
$|g_{K^*N\Lambda^*}|=11$ for the numerical calculation. In the
non-relativistic quark model, if $\Lambda^*$ is described as a
$p$--wave excitation of flavor-singlet spin-3/2 state, it is shown
that the strength of the $K^*N\Lambda^*$ coupling constant is of the
same order as that of $KN\Lambda^*$ or even larger than that. The
coupling constant of $g_{\gamma K^{*}K}$ is  
taken to be $0.254\,[{\rm GeV}^{-1}]$ for the charged decay and
$0.388\,[{\rm GeV}^{-1}]$ for the neutral 
 decay~\cite{Eidelman:2004wy}.

Taking all of these into consideration, we construct the invariant
amplitudes as follows : 
\bee
i\mathcal{M}_{s}&=&-\frac{eg_{KN\Lambda^*}}{M_{K}}\bar{u}^{\mu}(p_{2},s_{2})k_{2\mu}{\gamma}_{5} 
\frac{(\Slash{p}_{1}+M_{p})F_{c}+\Slash{k}_{1}F_{s}}{q^{2}_{s}-M^{2}_{p}}
\Slash{\epsilon}u(p_{1},s_{1}),\nn\\&+&\frac{e\kappa_{p}f_{KN\Lambda^*}}{2M_{p}M_{K}}
\bar{u}^{\mu}(p_{2},s_{2})k_{2\mu}{\gamma}_{5} 
\frac{(\Slash{q}_{s}+M_{p})F_{s}}{q^{2}_{s}-M^{2}_{p}}
\Slash{\epsilon}\Slash{k}_{1}u(p_{1},s_{1})\nn\\
i\mathcal{M}_{u}&=&-\frac{f_{KN\lambda}\kappa_{\Lambda^*}}{2M_{K}M_{\Lambda}}
\bar{u}_{\mu}(p_2)\Slash{k}_{1}\Slash{\epsilon}D^{\mu}_{\sigma}
\Theta^{\sigma\rho}k_{2\rho}\gamma_{5}u(p_1)F_{u},\nn\\ 
\mathcal{M}_{t}&=&\frac{2ef_{KN\Lambda^*}}{M_K}
\bar{u}^{\mu}(p_{2},s_{2})\frac{q_{t,\mu}k_{2}\cdot\epsilon}{q^{2}_{t}-M^{2}_{K}}
{\gamma}_{5}u(p_{1},s_{1})F_{c},\nn\\
i\mathcal{M}_{c}&=&\frac{ef_{KN\Lambda^*}}{M_K}\bar{u}^{\mu}(p_{2},s_{2})
\epsilon_{\mu}{\gamma}_{5}u(p_{1},s_{1})F_{c},\nn\\i\mathcal{M}_{v}&=&
\frac{-ig_{\gamma{K}K^*}g_{K^{*}NB}}{M_{K^{*}}(q^{2}_{t}-M^{2}_{K})}   
\bar{u}^{\mu}(p_{2},s_{2})\gamma_{\nu}\left(q^{\mu}_{t}g^{\nu\sigma}-
g^{\nu}_{t}q^{\mu\sigma}\right)\epsilon_{\rho\eta\xi\sigma}k^{\rho}_{1}
\epsilon^{\eta}k^{\xi}_{2}u(p_{1},s_{1})F_{v},\nn\\
\label{amplitudes}  
\eee
where ${u}^{\mu}$ is the RS vector-spinor  which
is defined as follows :
\bee
u^{\mu}(p_{2},s_2)=\sum_{\lambda,s}\left(1\lambda\frac{1}{2}s|
\frac{3}{2}s_{2}\right)e^{\mu}(p_{2},\lambda)u(p_{2},s),
\eee 
with the Clebsh-Gordon coefficient
$(1\lambda\frac{1}{2}s|\frac{3}{2}s_{2})$.  $D_{\mu\nu}$ stands for
the spin-3/2 propagator : 
\bee
D_{\mu\nu}=-\frac{\Slash{q}+M_{\Lambda^*}}{q^2-M^2_{\Lambda^*}}\left\{g_{\mu\nu}-\frac{1}{3}\gamma_{\mu}\gamma_{\nu}-
\frac{2}{3M^{2}_{\Lambda^*}}q_{\mu}q_{\nu}+\frac{q_{\mu}\gamma_{\nu}+
q_{\mu}\gamma_{\nu}}{3M_{\Lambda^*}}\right\}.
\eee
In Eq.~(\ref{amplitudes}), we have shown how the four-dimensional form
factor is inserted in such way that gauge-invariance is preserved. As
suggested in Ref.~\cite{Haberzettl:1998eq,Davidson:2001qs}, we
adopt the following parameterization 
for the four-dimensional gauge-- and Lorentz--invariant form factor:
\bee
F_{x}(q^2)&=&\frac{\Lambda^4}{\Lambda^4+(x-M^2_x)^2},\,\,x=s,t,u,v\nn\\
F_{c}&=&F_{s}+F_{t}-F_{s}F_{t}. 
\label{formfactor1}
\eee
The form of $F_{c}$ is chosen such that the on-shell values of the
coupling constants are reproduced. The cutoff
masses, $\Lambda$ will be determined to produce the
$\gamma p\to K^+\Lambda^*$ total cross section data. 
\section{Numerical results}
\subsection{$\gamma N\to K\Lambda^*$ without the form factors}
In this subsection, we present the numerical results for the $\gamma N\to
K\Lambda^*$ without the form factors in order to see the  
bare contributions from each channel.
\begin{figure}[tbh]
\begin{tabular}{cc}
\includegraphics[width=6cm]{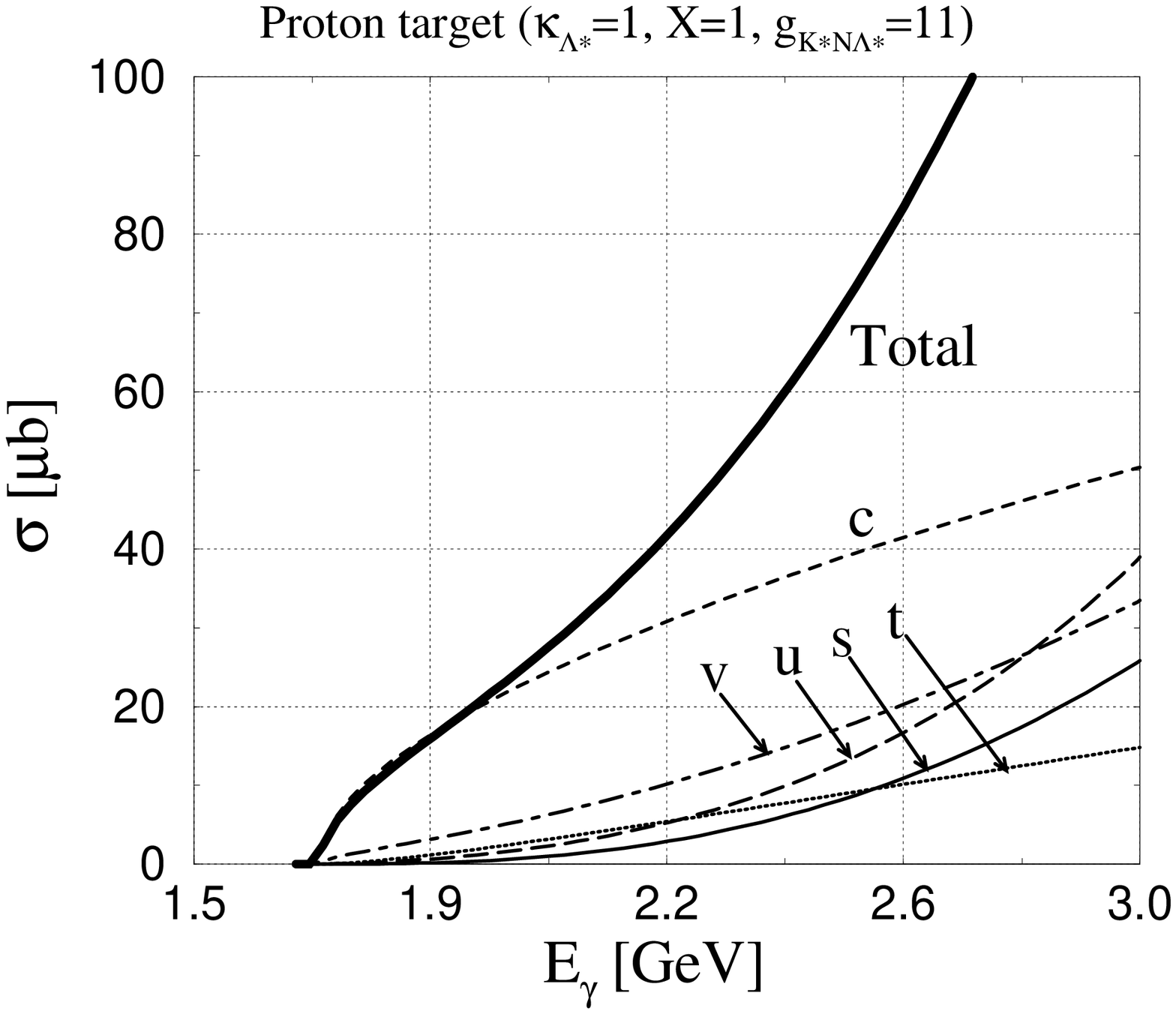}
\includegraphics[width=6cm]{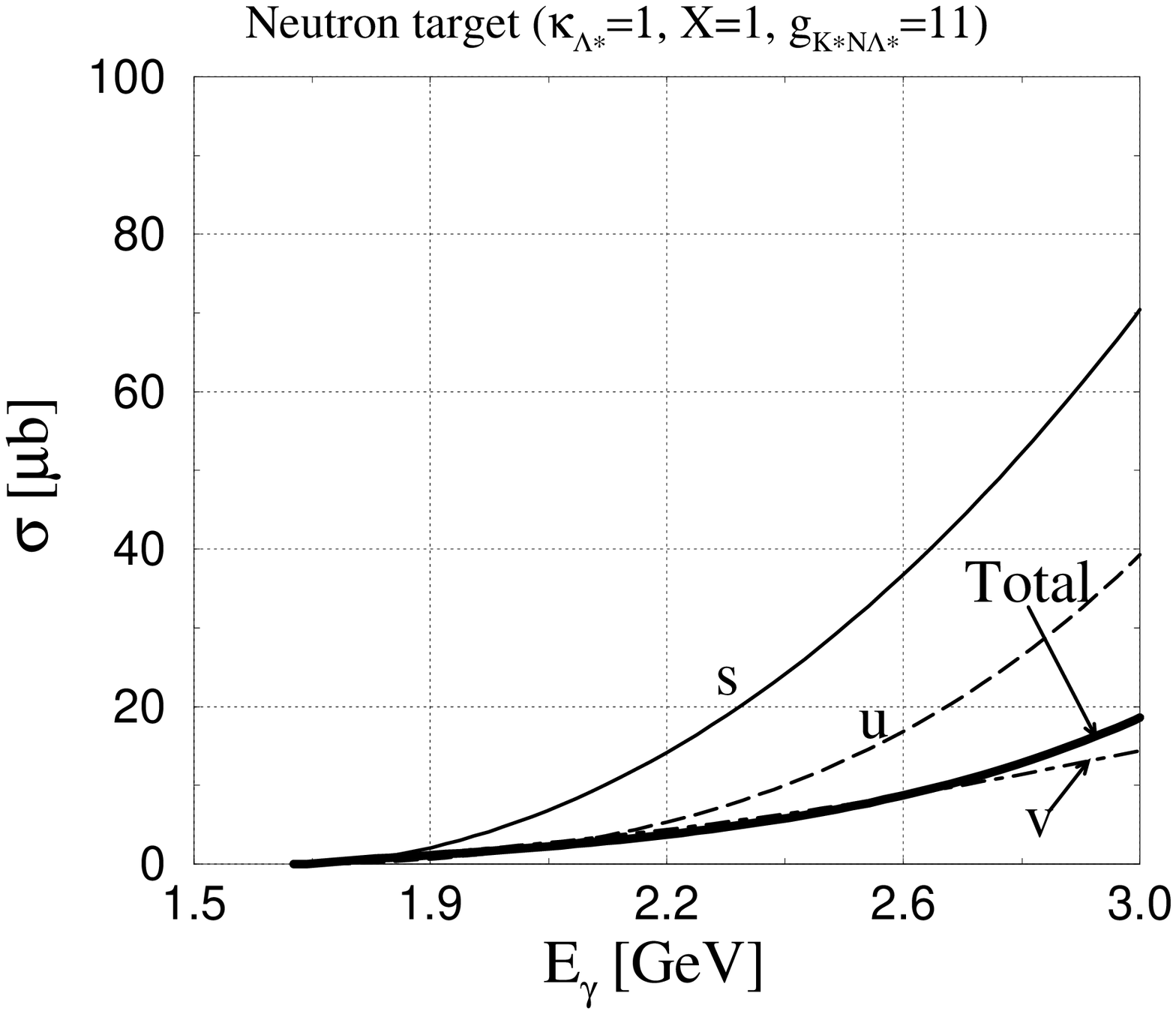}
\end{tabular}
\begin{tabular}{cc}
\includegraphics[width=6cm]{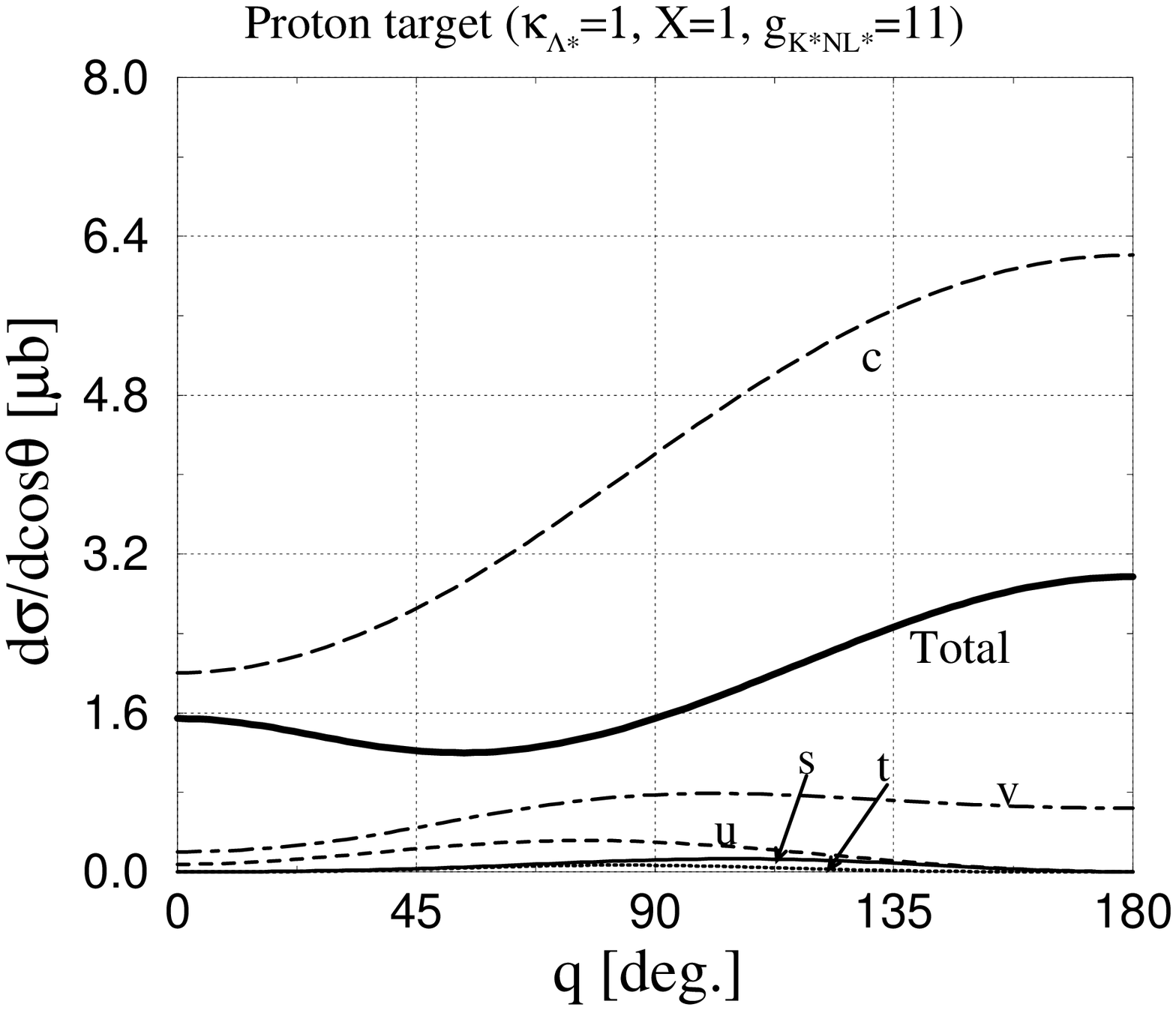}
\includegraphics[width=6cm]{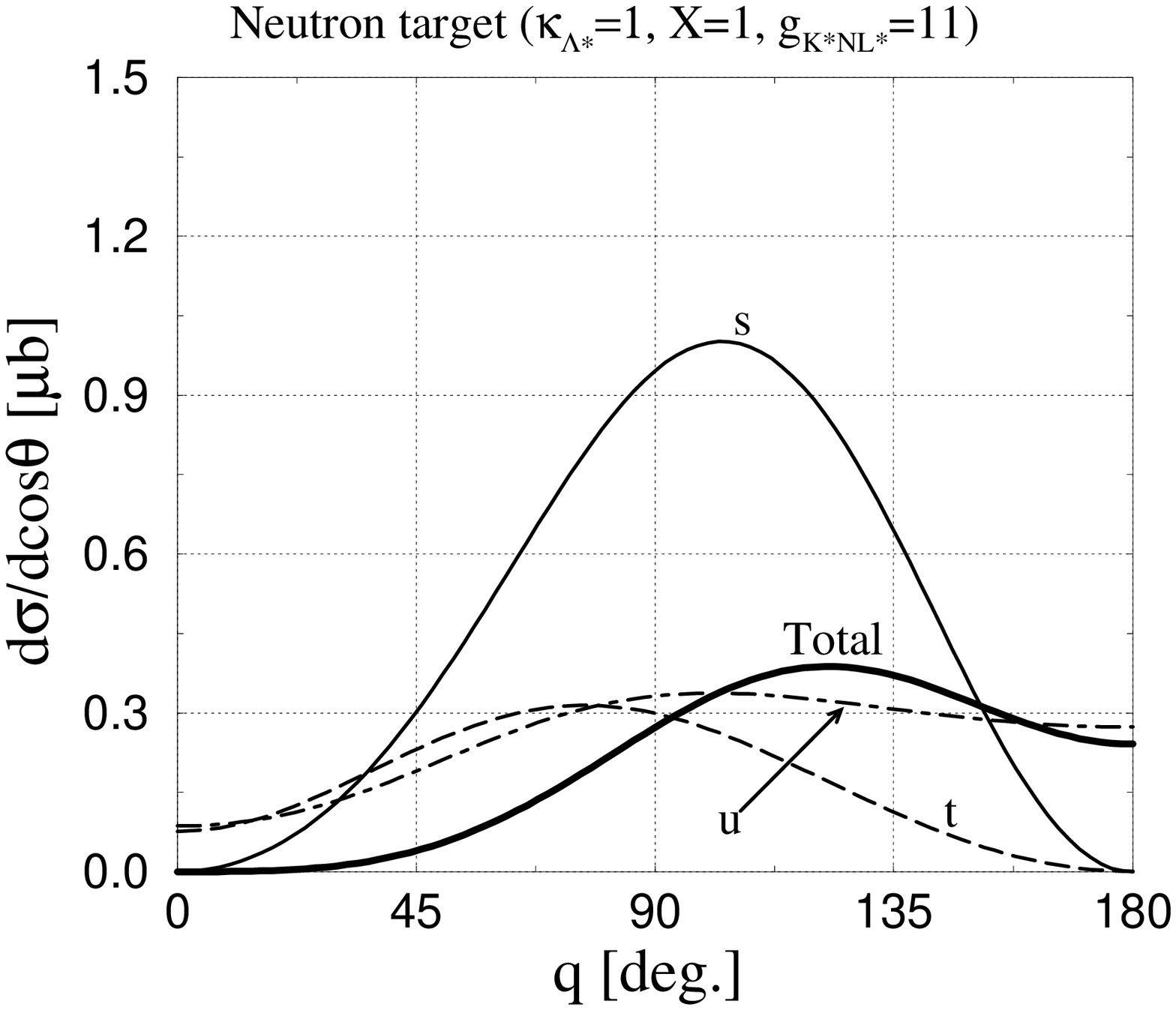}
\end{tabular}
\caption{Various contributions to the total cross (upper two panels)
and differential (lower two panels) sections for the
proton target (left panel) and 
the neutron target (right panel) without a form factor. The
differential cross sections are obtained for $E_{\gamma}=2..0$ GeV. We choose the
unknown parameters $(\kappa_{\Lambda^*},X)=(1,1)$ and the coupling
constant $g_{KN\Lambda^*}=+11$.}
\label{fig1}
\end{figure} 
In the upper two panels of Fig.~\ref{fig1}, we present various
contributions to the total 
cross sections for the proton 
and neutron targets without form factors. Here, we fix the
unknown parameters $\kappa_{\Lambda^*}=1.0$, $X=1$ 
and $g_{K^*N\Lambda^*}=g_{KN\Lambda^*}=+11$. Later, we discuss
parameter dependence of our calculation. In the case of the proton
target in the 
left panel of Fig.~\ref{fig1}, we  
observe that the contribution of the contact term  
($c$--channel) is dominant over other channels. We also note that the
$c$-- and 
$v$--channels demonstrate the $s$--wave threshold 
behavior ($\sigma\sim (E_{\gamma}-E_{\rm th})^{1/2}$), where $E_{\rm
  th}$ stands for the threshold energy, although the magnitude 
of the $v$--channel is smaller than that of the $c$--channel. However,
the other channels, $s$, $u$ and $t$ start with $p$-wave contribution
in the vicinity of the threshold. This kinematical effect explains why the
$s$--, $u$-- and $t$--channels are relatively suppressed especially
near the threshold region as compared
with the $c$-- and $v$--channels containing the $s$-wave
contribution.  
For the
neutron target (right panel), the $s$--channel is the dominant
contribution, because we 
have no neutral charge coupling in the $c$--channel. We also find
destructive interference between the $s$--, $u$-- and
$v$--channels. We also show the differential cross sections in the
upper two panels of Fig.~\ref{fig1} without the form factor. The
differential cross sections are obtained for $E_{\gamma}=2..0$ GeV.  
\subsection{$\gamma N\to K\Lambda^*$ with the form factor}
In this subsection, we present the numerical results for the $\Lambda^*$
photoproduction with the form factor given in 
Eq.~(\ref{formfactor1}). The experimental data are taken from
Ref.~\cite{Barber:1980zv} 
\begin{figure}[tbh]
\begin{center}
\includegraphics[width=6cm]{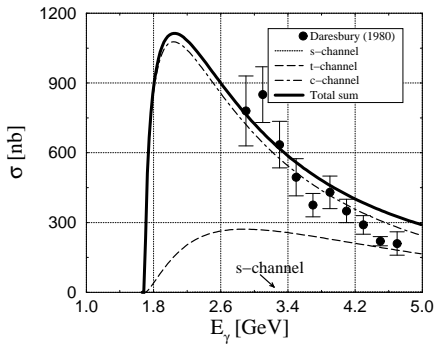}
\end{center}
\caption{The total cross sections for the proton target with the
form factor. We show the $s$--, $t$-- and
$c$--channels, separately. The contribution of the $s$--channel is
negligibly small.}
\label{fig2}
\end{figure} 
In Fig.~\ref{fig2}, we show the total cross sections including $s$--,
$t$-- and $c$--channels which do not contain unknown parameters
$\kappa_{\Lambda^*}$, $X$ and $g_{K^*N\Lambda^*}$. The experimental
data are  taken from Ref.~\cite{Barber:1980zv} with  
photon energies in the range of $2.8$ GeV $<\, E_{\gamma}\,<$  $4.8$ GeV. In
the present calculation, we have chosen the cutoff parameter
$\Lambda=700$ MeV, which can reproduce the experimental data very
well from the threshold to the higher energy region. However, the
results in the higher energy region should not be taken seriously  
since the Born approximation works properly in the low energy region near
the threshold. In fact, we have verified that the total cross sections
depend much on the parameters, $\kappa_{\Lambda^*}$ and $X$, beyond
$E_{\gamma}\gsim3$ GeV, whereas the parameter dependence is rather weak for
$E_{\gamma}\lsim3$ GeV~\cite{nam1}. Therefore, we focus most of our
discussion below the energy region $E_{\gamma}\lsim3$ GeV, where the
Born approximation of the effective Lagrangian method is expected to
work. It is interesting to observe that the size and energy dependence
of the total cross section of the $\Lambda^*(1520)$ production
are similar to those of the production of the ground state
$\Lambda(1116)$~\cite{Boyarski:1970yc,Barber:1980zv}.

In Fig.~\ref{fig2}, we have also shown separate contributions from the
$s$--, $t$-- and $c$--channels. Obviously, the $c$--channels
contribution dominates, which is the important feature of the reaction
with the proton. In order to see the role of
$K^*$--exchange, we consider the
two cases $g_{K^*N\Lambda^*}=\pm |g_{KN\Lambda^*}|$ and compare the
results to the case without the $K^*N\Lambda^*$ coupling,
i.e. $g_{K^*N\Lambda^*}=0$.   
\begin{figure}[tbh]
\begin{tabular}{cc}
\includegraphics[width=6cm]{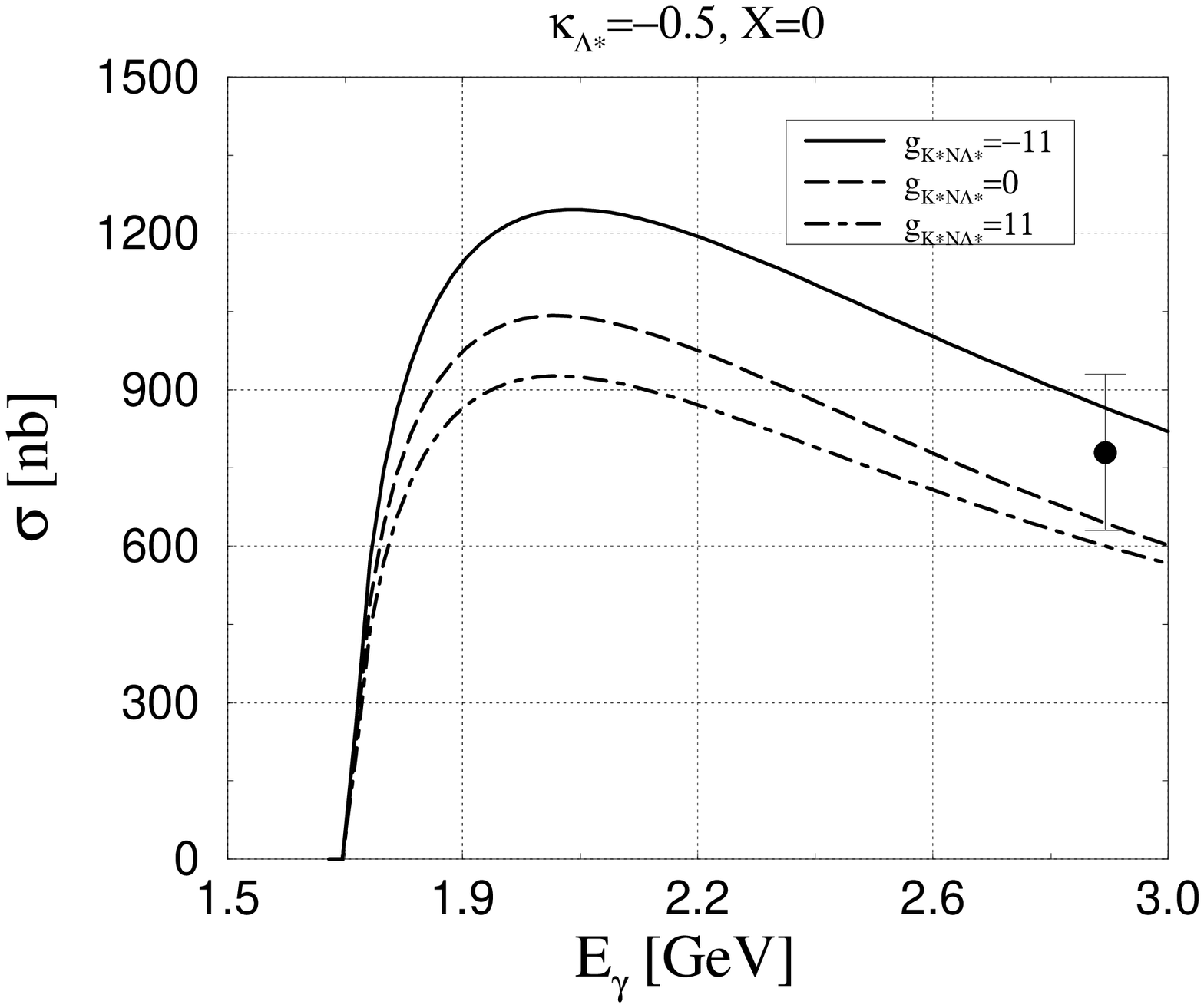}
\includegraphics[width=6cm]{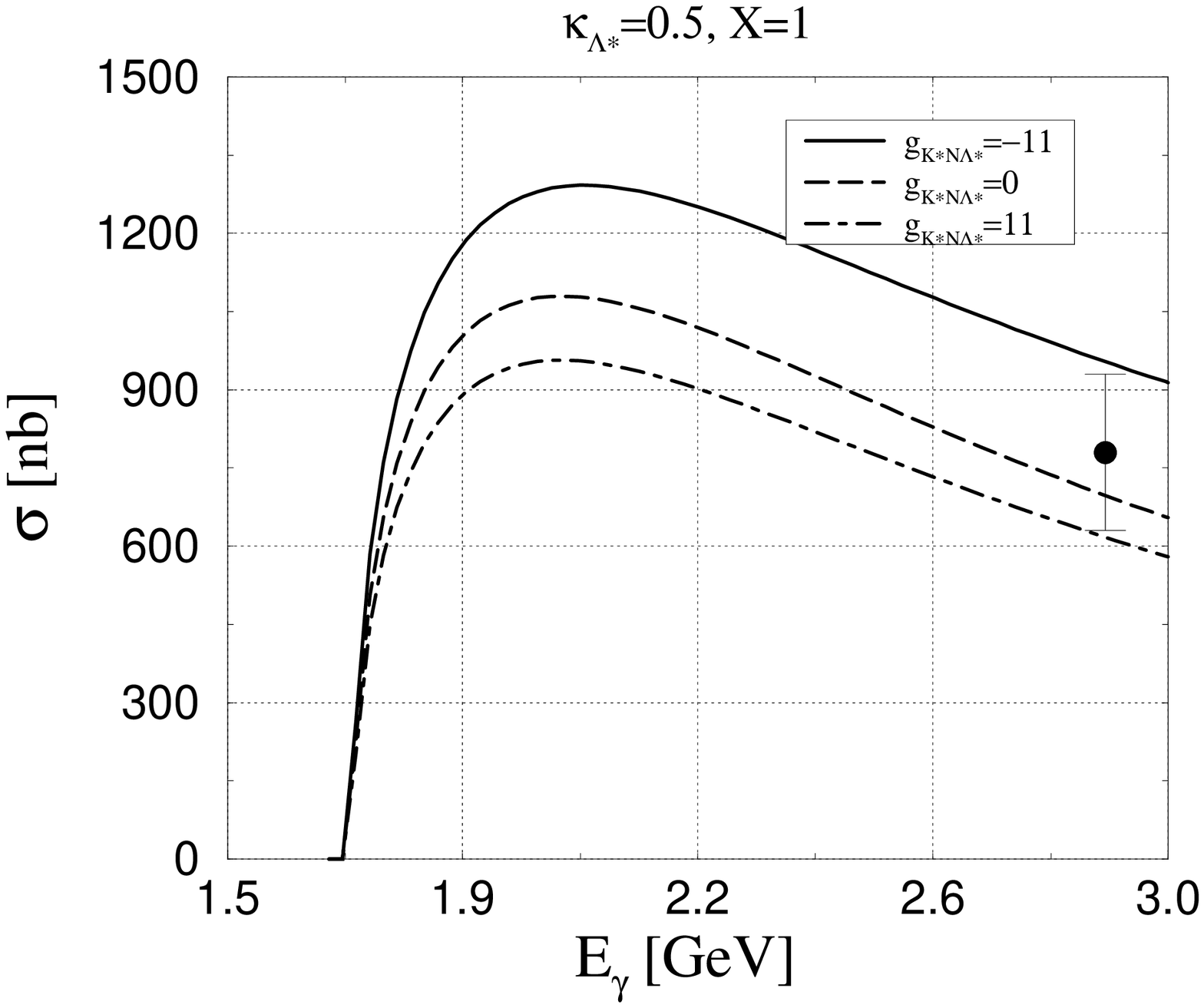}
\end{tabular}
\caption{The total cross sections for the proton target with the
form factor. We choose $(\kappa_{\Lambda^*},X) = (-0.5,0)$ and
$(0.5,1)$ in order to see the parameter dependence. We choose three
different values of the 
coupling constant $g_{K^*N\Lambda^*}=0$ and $\pm11$.} 
\label{fig3}
\end{figure} 
In Fig.~\ref{fig3}, we compare the total cross sections up to
$E_{\gamma}\lsim3$ GeV using the different $K^*N\Lambda^*$ coupling
strengths. The total cross sections are rather insensitive to the
contribution of $K^*$--exchange, a consequence that the the $c$--channel 
contribution is still dominant for the proton target. 

These two results are compared for the two different parameter sets,
$(\kappa_{\Lambda^*},X)=(-0.5,0)$ 
and $(0.5,1)$. As discussed previously, the results do not depend
much on these parameters at low-energies $E_{\gamma}\lsim3$ GeV. In
the quark model, it is found that the anomalous magnetic moment
$\kappa_{\Lambda^*}$ turns out to vanish in the $SU(3)$
limit for a pure flavor singlet $\Lambda^*$. Taking into account that
the effect of explicit $SU(3)$ 
symmetry breaking, we expect that the values of $\kappa_{\Lambda^*}$
may lie in the range of $|\kappa_{\Lambda^*}|<0.5$. However, from
Fig.~\ref{fig3}, we expect that the dependence on $\kappa_{\Lambda^*}$
within this range is small. Therefore, these two parameters
$\kappa_{\Lambda^*}$ and $X$ can be set zero, i.e. $\kappa_{\Lambda^*}=X=0$.

\begin{figure}[tbh]
\begin{center}
\includegraphics[width=6cm]{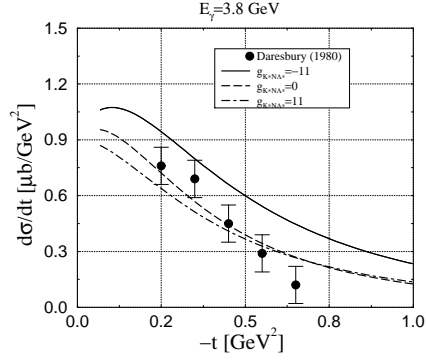}
\end{center}
\caption{The $t$--dependence for the proton target at
$E_{\gamma}=3.8$ GeV. We choose  $(\kappa_{\Lambda^*},X) = (0,0)$.}  
\label{fig4}
\end{figure} 

\begin{figure}[tbh]
\begin{tabular}{ccc}
\includegraphics[width=4cm]{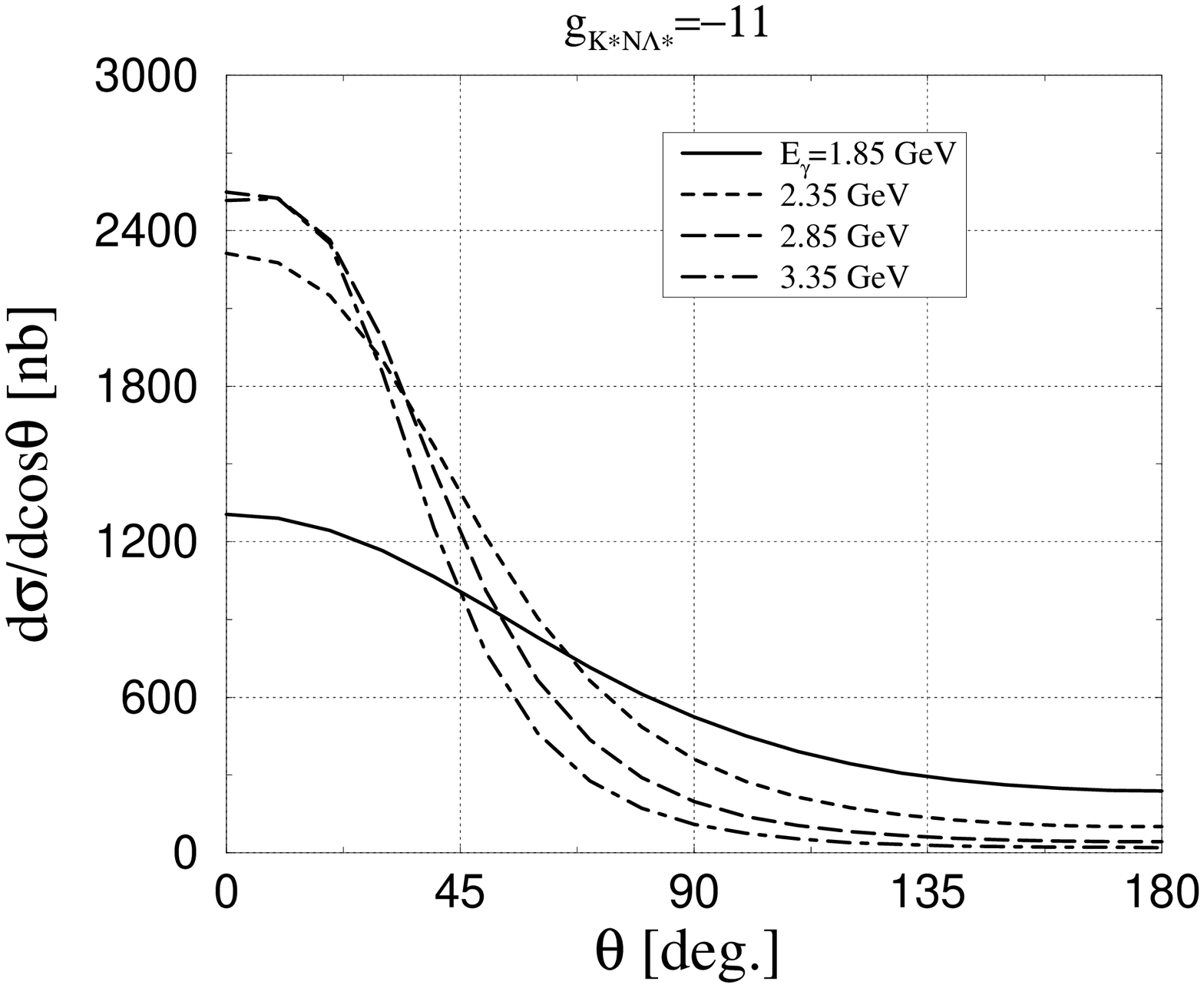}
\includegraphics[width=4cm]{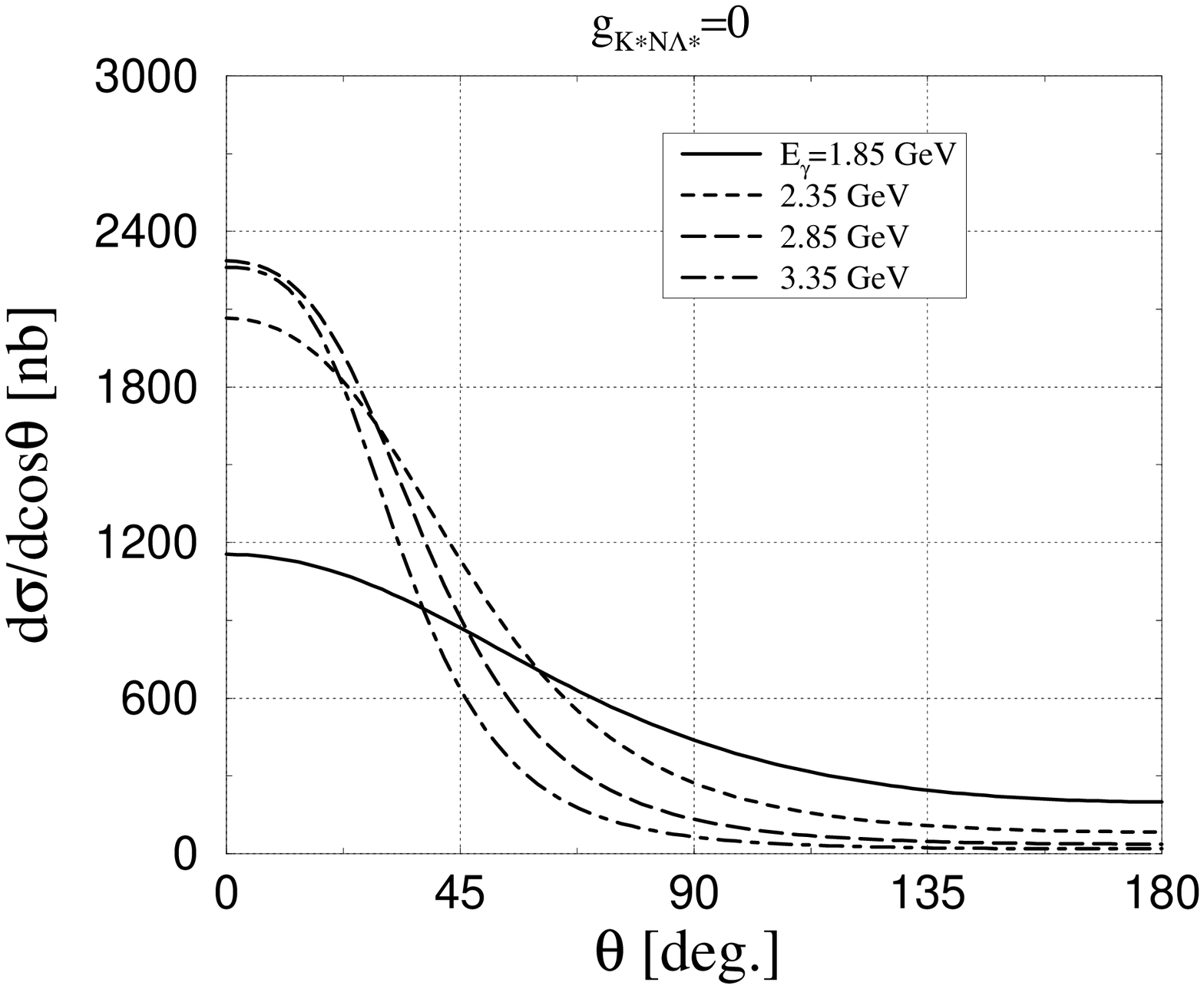}
\includegraphics[width=4cm]{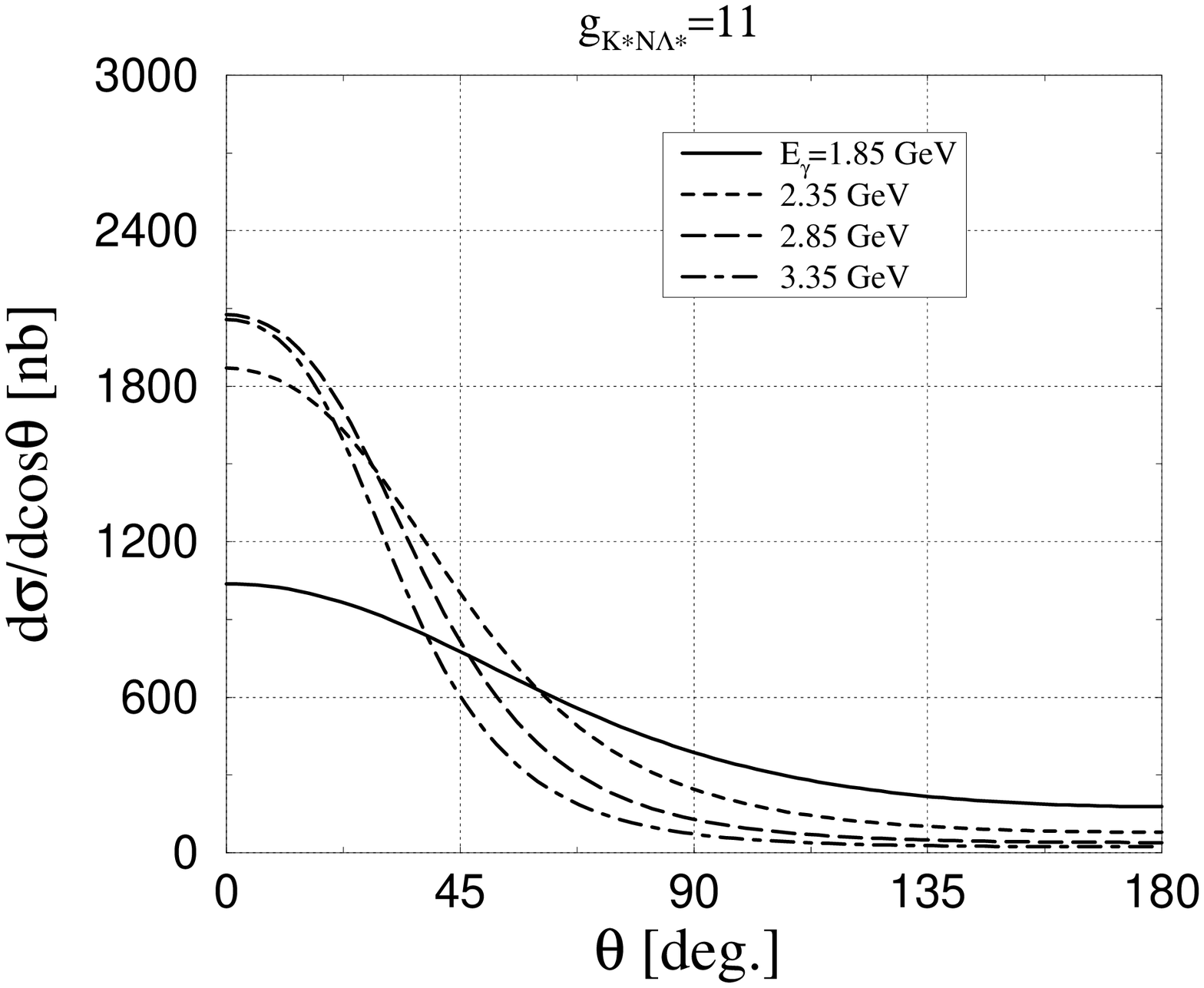}
\end{tabular}
\caption{The differential cross sections for the proton target with the
form factor.  Several photon energies are taken into
account. We choose  $(\kappa_{\Lambda^*},X) = (0,0)$.}  
\label{fig5}
\end{figure}

In Fig.~\ref{fig4}, we plot the dependence on the momentum transfer,
${d\sigma}/{dt}$ ($t$--dependence)  at 
$E_{\gamma}=3.8$ GeV which is the average energy of the Daresbury
experiment  ($2.8<E_{\gamma}<4.8$ GeV)~\cite{Barber:1980zv}. The
figure indicates good agreement with the data. In 
Fig.~\ref{fig5}, we also demonstrate the angular dependence. Here, 
$\theta$ is the angle between the incident photon and the outgoing
kaon in the center of mass system. Each panel draws the differential
cross sections ${d\sigma}/{d(\cos\theta)}$ with $g_{K^*N\Lambda^*}$
varied. We observe that $K^*$--exchange does not contribute much to
the differential cross sections as in the case of the total cross
sections (see Fig.~\ref{fig3}).

\begin{figure}[tbh]
\begin{tabular}{cc}
\includegraphics[width=6cm]{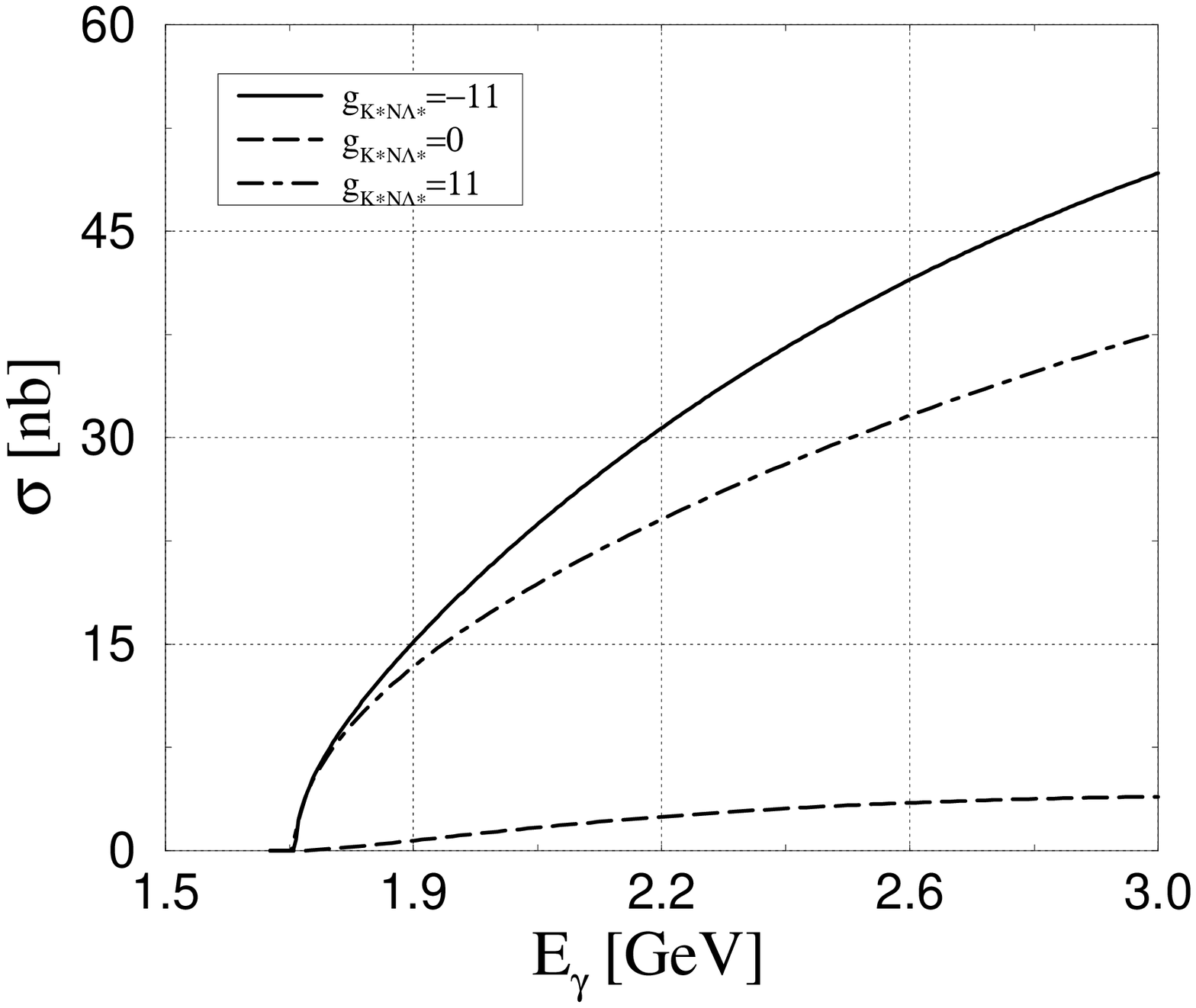}
\includegraphics[width=6cm]{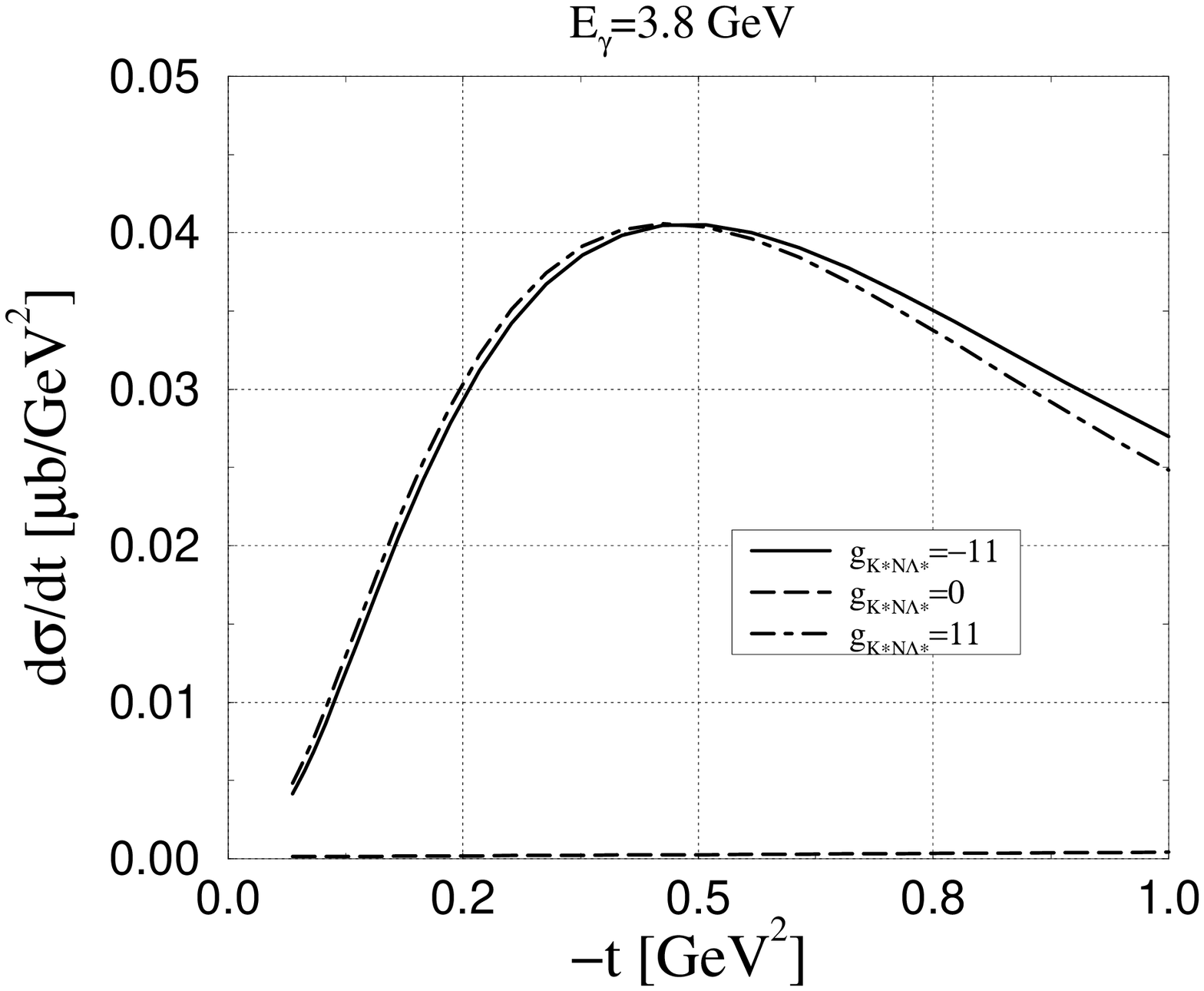}
\end{tabular}
\caption{The left panel : the total cross sections for the neutron target with the
form factor. We use $(\kappa_{\Lambda^*},X) = (0,0)$. We choose three different values of the
coupling constants $g_{K^*N\Lambda^*}=0$ and $\pm 11$. The right
panel : the $t$--dependence for the neutron target
with the form factor at $E_{\gamma}=3.8$ GeV.} 
\label{fig6}
\end{figure}

Now, we discuss the case of the neutron target. In this case, the
$c$--channel contribution vanishes, and therefore, the total cross
section becomes much smaller than the case of the proton target
especially when
the form factor is employed. The left panel of Fig.~\ref{fig6}
shows the total cross sections with and without $K^*$--exchange. When
$K^*$--exchange is switched off, only the $s$--channel plays a role.
Furthermore, in the neutron target case, the $p$--wave contribution is
dominant, which leads to the energy dependence $(E_{\gamma}-E_{\rm
  th})^{3/2}$ of the total cross sections  
as clearly indicated in Fig.~\ref{fig6}. 
The inclusion of $K^*$--exchange enhances
the total cross sections with the energy dependence 
$\sim(E_{\gamma}-E_{\rm th})^{1/2}$. The contribution of $K^*$--exchange is important for the neutron
target. However, the magnitude of that contribution is still
small as compared to the proton target. Experimental study of the
energy dependence will be useful to obtain the informations of the
reaction mechanism.

Turning to the $t$--dependence,
we demonstrate it in the right panel of Fig.~\ref{fig6}, where we choose once again 
$E_{\gamma}=3.8$ GeV. The $t$--dependence of the neutron target is
very much different from the proton one, since the relevant diagrams
are different.       

\begin{figure}[tbh]
\begin{tabular}{ccc}
\includegraphics[width=4cm]{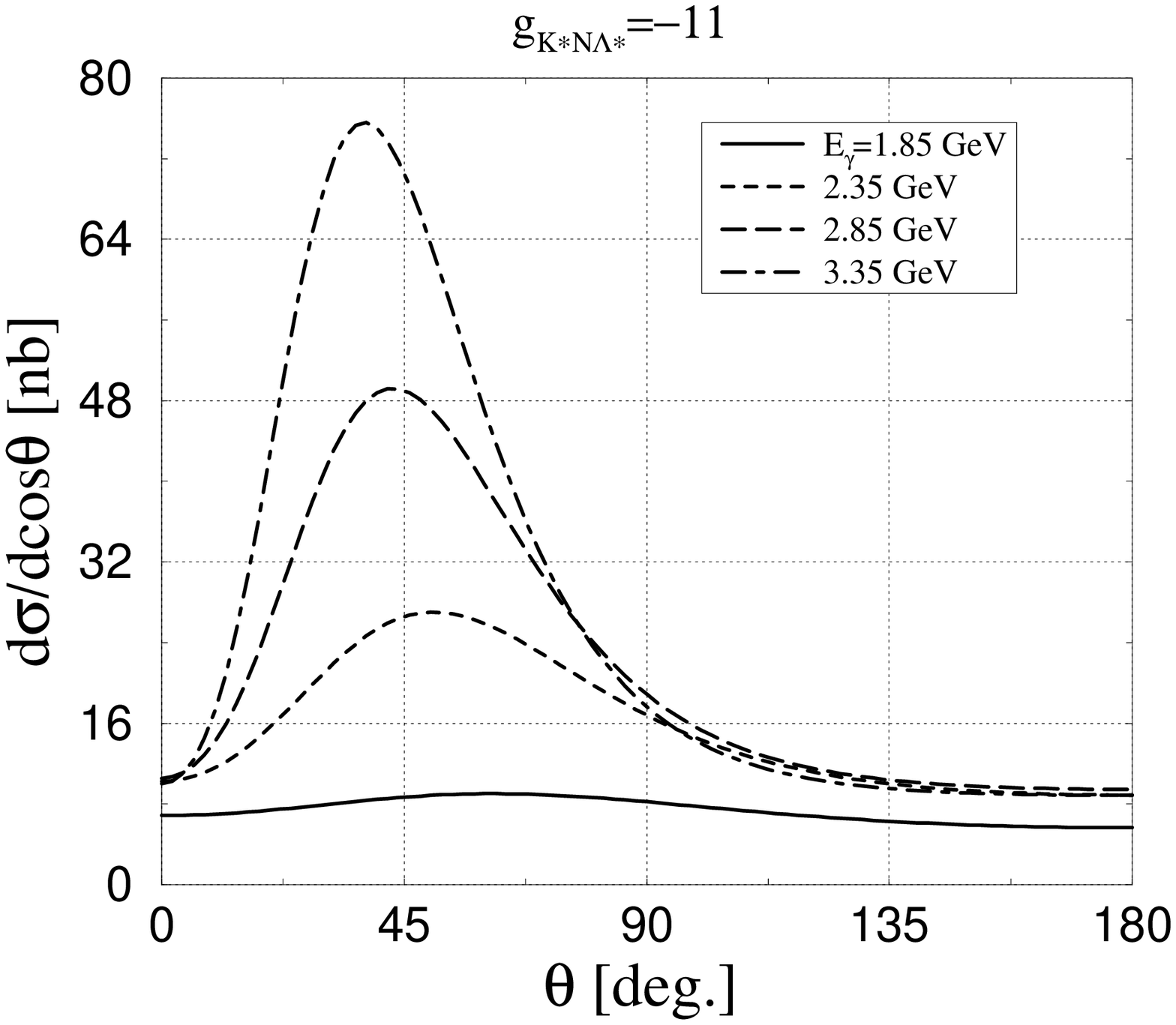}
\includegraphics[width=4cm]{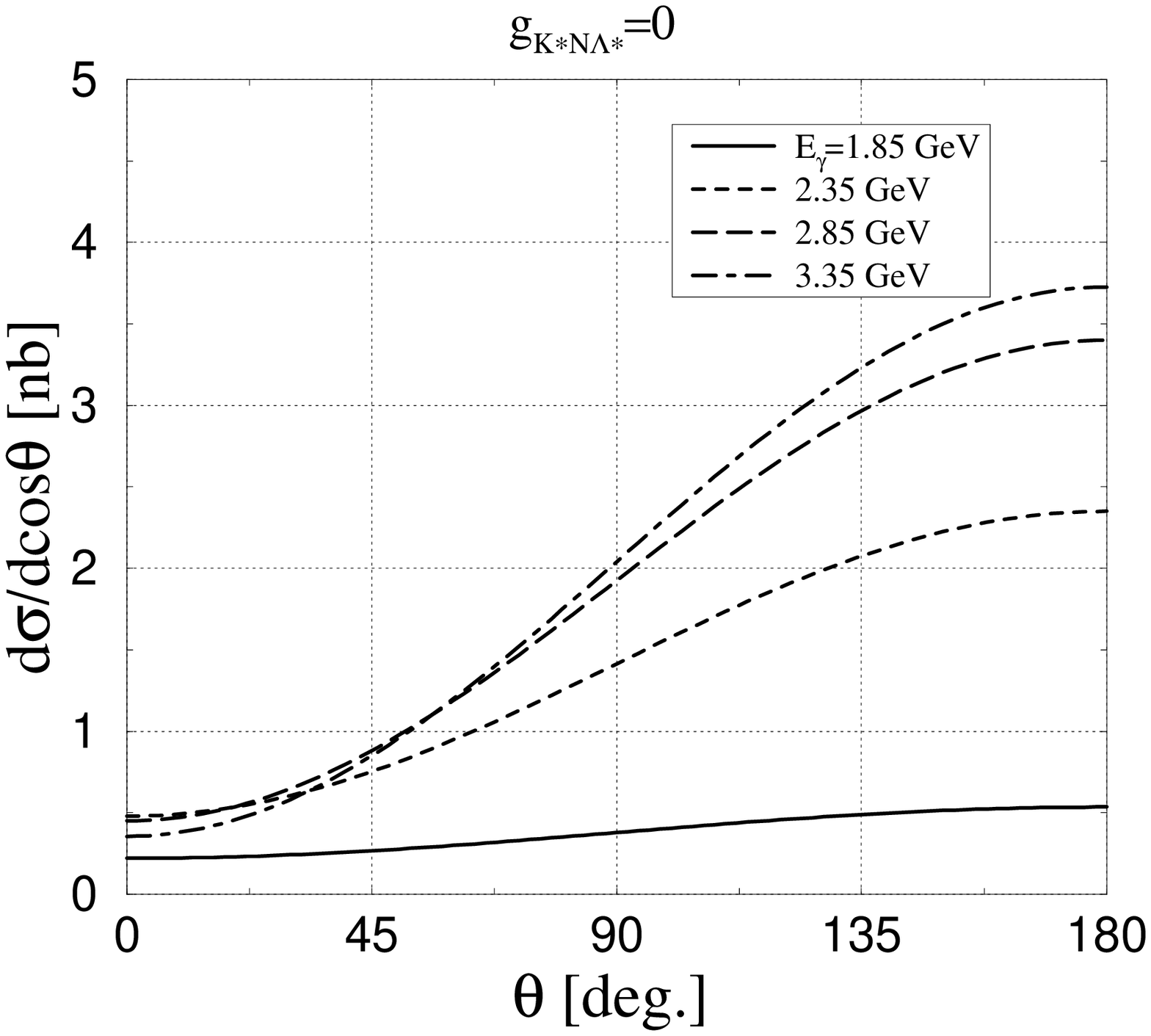}
\includegraphics[width=4cm]{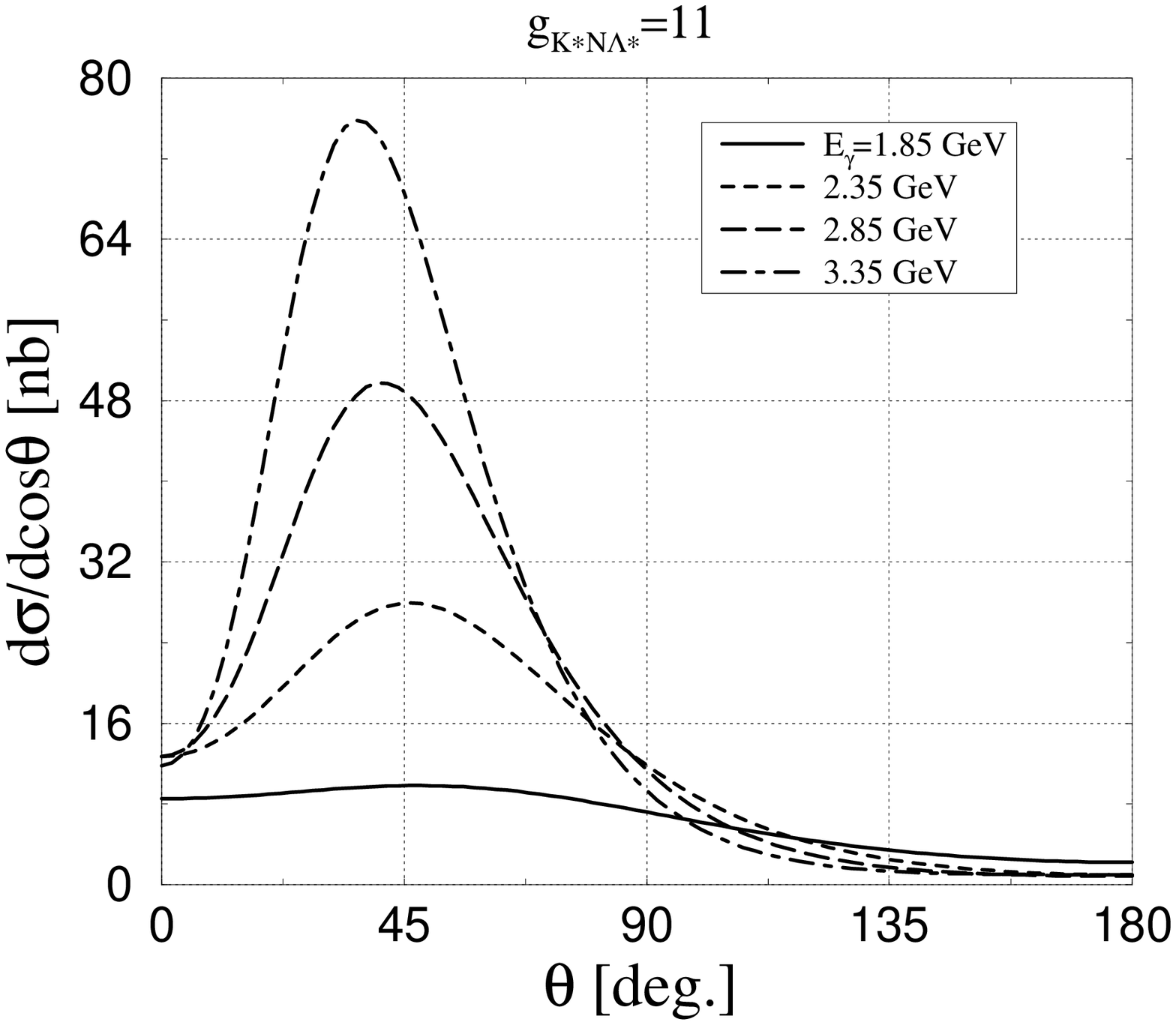}
\end{tabular}
\caption{The differential cross sections for the neutron target with the
form factor.  Several photon energies are taken into
account. We choose  $(\kappa_{\Lambda^*},X) = (0,0)$.}
\label{fig7}
\end{figure}

In Fig.~\ref{fig7}, we show the angular dependence for the neutron
target, using the form factor. With the $K^*$--exchange being included,
the differential cross sections are enhanced around $\sim
45^{\circ}$. Note that the sign of $g_{K^*N\Lambda^*}$ is not
important. The bump around $45^{\circ}$ is a typical behavior comming
from the $K^*$--exchange.  

\section{The role of $K^*$--exchange}
In Ref.~\cite{Barber:1980zv} of the Daresbury experiment, it was
argued that 
the $\Lambda^*$ photoproduction 
was dominated by vector $K^*$--exchange ($v$--channel) rather than 
pseudoscalar $K$--exchange ($t$--channel) by analyzing the decay
amplitude in the $t$--channel in the helicity basis of the 
$\Lambda^*$. If the helicity of the $\Lambda^*$ is
$S_z=\pm3/2$, the decay of $\Lambda^*\to K^-p$ is explained by
$\sin^2\theta$ in which $\theta$ is the angle between
the two kaons in the helicity basis (see Ref.~\cite{Barrow:2001ds} for
details).  On the other hand,  the
angular dependence becomes $1/3+\cos^2\theta$ for the decay of the
$S_z=\pm1/2$ state. Therefore, 
taking into account the ratio of these two helicity amplitudes, one
could extract information which meson would dominate.  In
Ref.~\cite{Barber:1980zv}, it was shown that the ratio of  
$(S_z=\pm1)/(S_z=\pm3/2)$ was nearly zero.  Thus, it was
suggested that the $\Lambda^*$ photoproduction was dominated by the
$v$--channel.  

\begin{figure}[tbh]
\begin{tabular}{ccc}
\includegraphics[width=4cm]{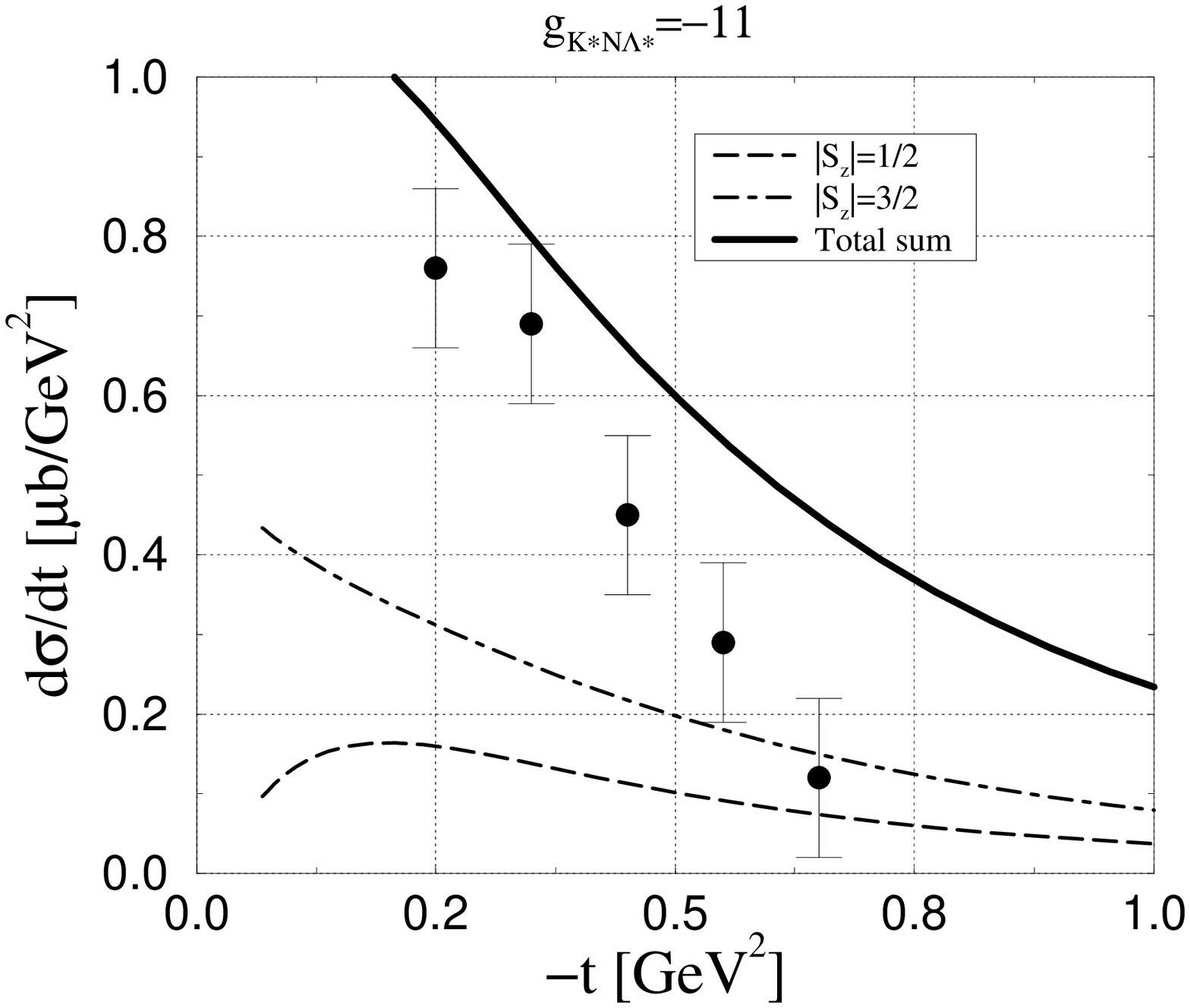}
\includegraphics[width=4cm]{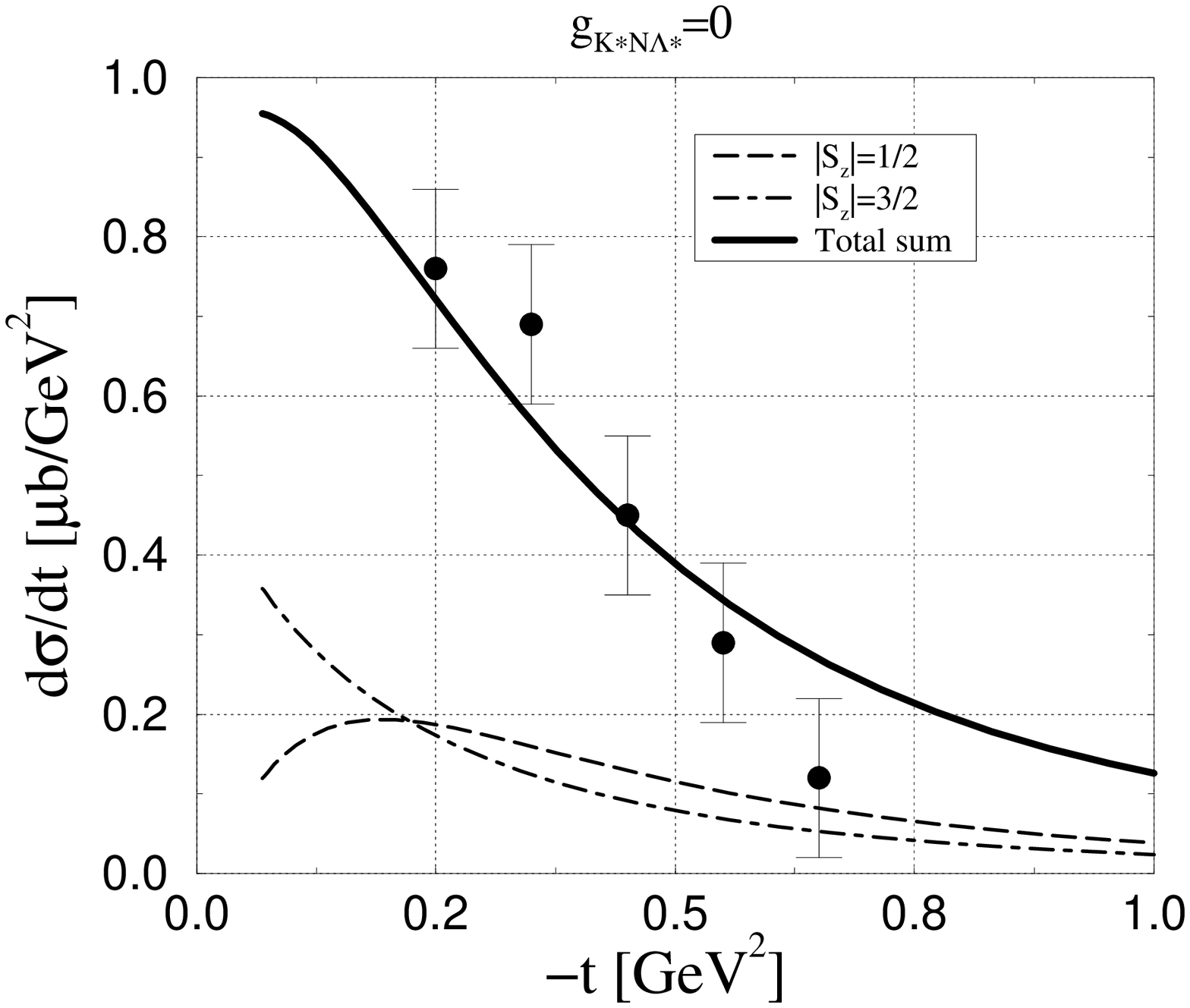}
\includegraphics[width=4cm]{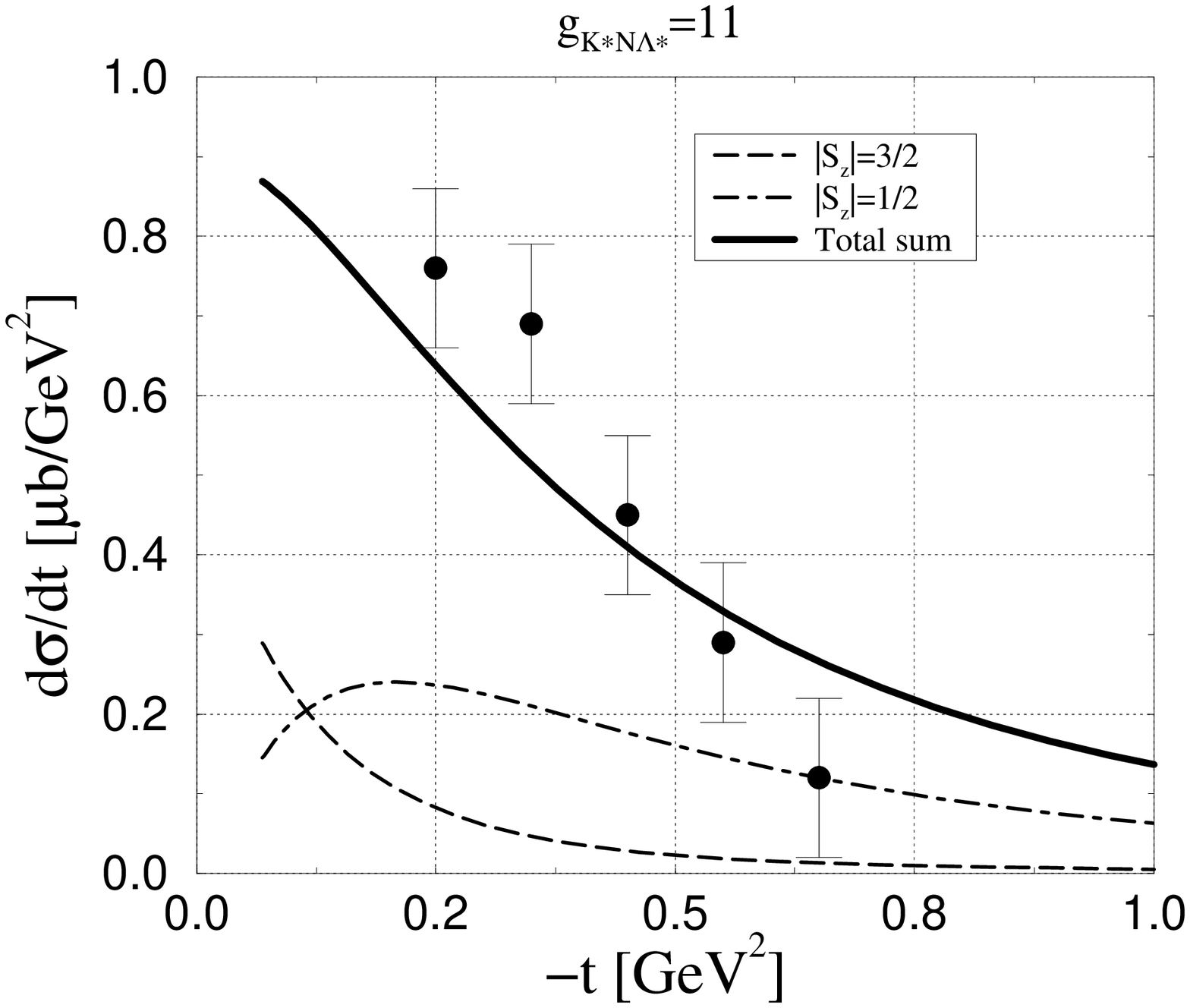}
\end{tabular}
\caption{The $t$--dependence for each helicity of the $\Lambda^*$ in
the final state.  We change the coupling 
constant $g_{K^*N\Lambda^*}$. We choose  $(\kappa_{\Lambda^*},X) = (0,0)$.}
\label{fig12}
\end{figure}

In Fig.~\ref{fig12}, we plot the $t$--dependence for each helicity using
the form factor with three different values for the coupling
constants $g_{K^*N\Lambda^*}$. We choose
$E_{\gamma}=3.8$ GeV as done previously.     
In Fig.~\ref{fig12}, we observe that the $S_z=\pm3/2$ contribution is 
dominant especially in the region $-t\lsim0.2\,{\rm GeV}^{-2}$.  There
is also a small contribution from the $S_z=\pm1/2$.  However, we find
that even without the $v$--channel ($g_{K^*N\Lambda^*}=0$),
the $S_z=\pm3/2$ does not become zero.  Therefore, the $S_z=\pm3/2$
contribution comes from not only the $v$--channel but also from the
other channels.  

\begin{figure}[tbh]
\begin{tabular}{ccc}
\includegraphics[width=4cm]{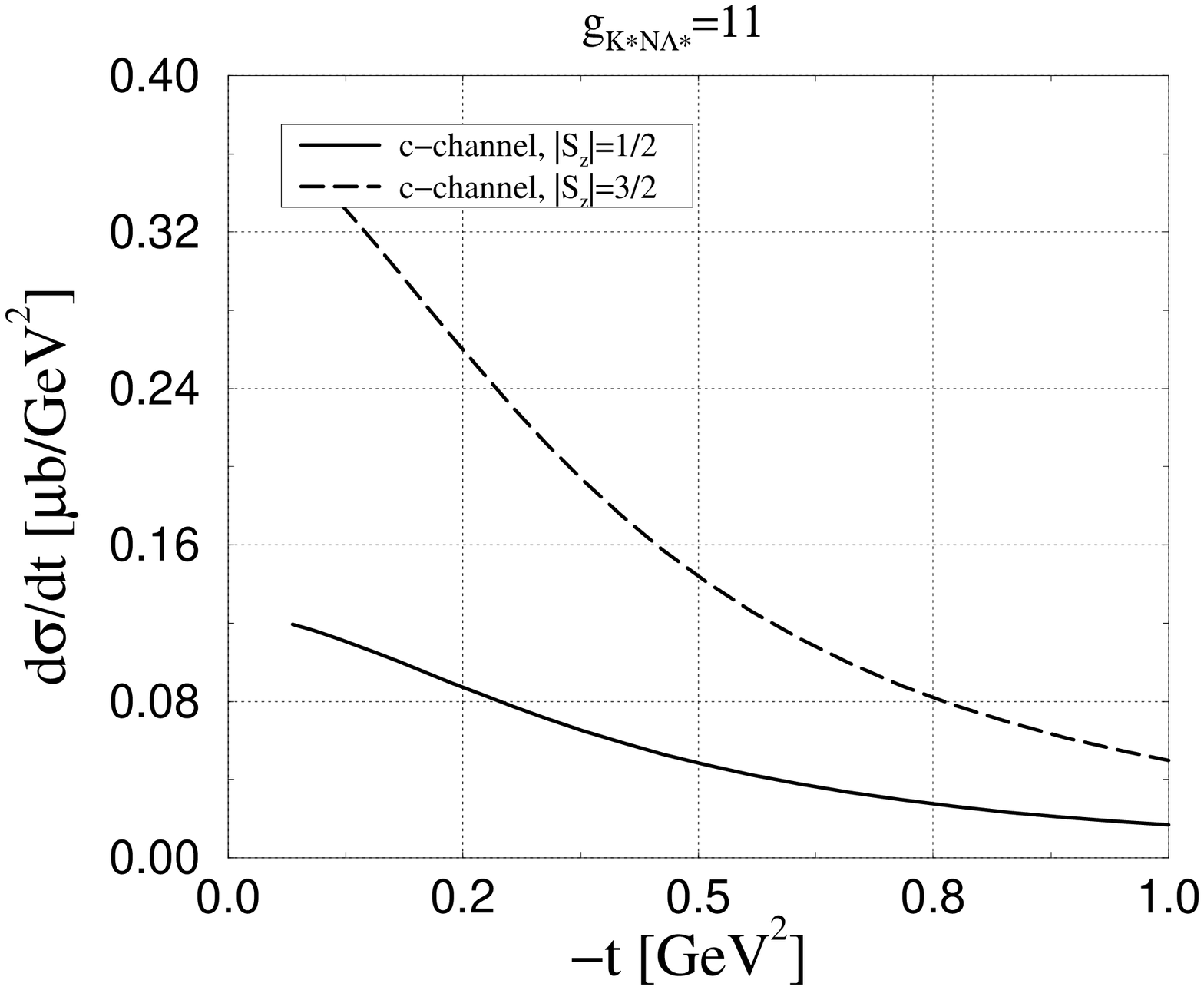}
\includegraphics[width=4cm]{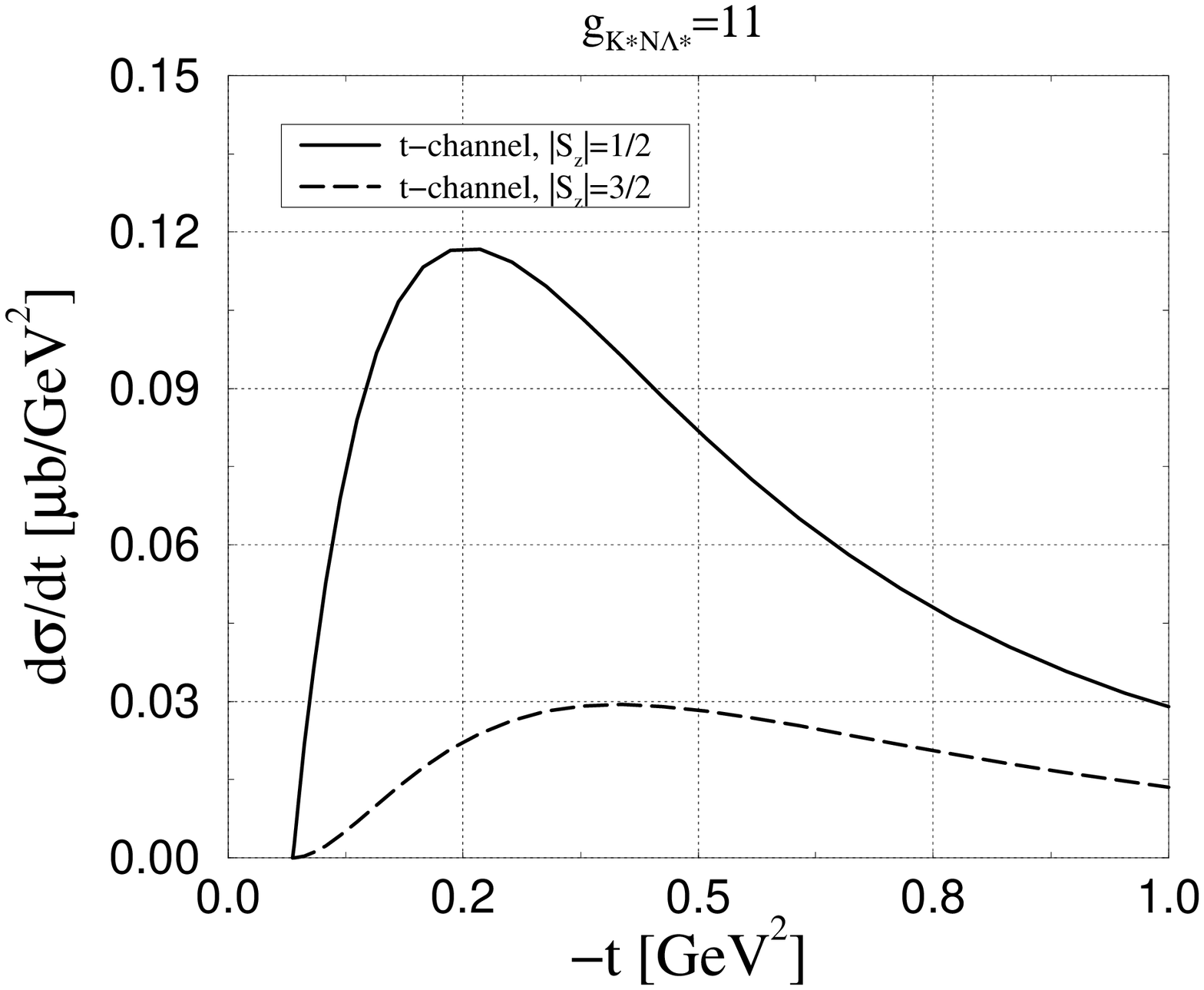}
\includegraphics[width=4cm]{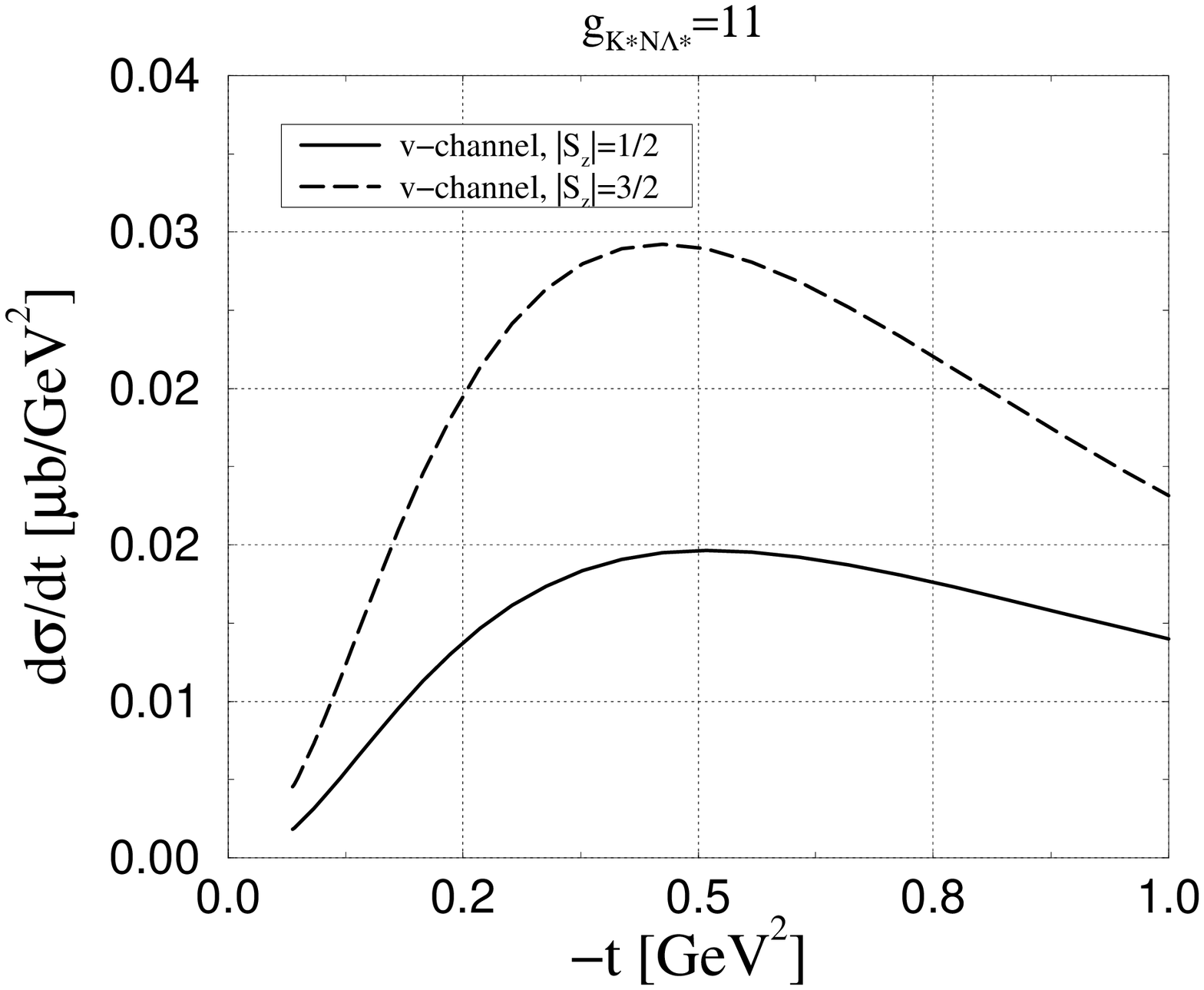}
\end{tabular}
\caption{The $t$--dependence for the two helicities $S_z=\pm 1/2$ and $S_z=\pm 3/2$ for the $c$--,
  $t$-- and $v$--channels.}
\label{fig13}
\end{figure}

In order to see this situation more 
carefully, we pick up three important channels, the $c$--,
$t$-- and $v$--channels, and plot the $t$--dependence for each
helicity in Fig.~\ref{fig13}.  One can see that the $S_z=\pm1/2$
contribution is larger than that of the  $S_z=\pm3/2$ for the
pseudoscalar $K$--exchange ($t$--channel), and vice versa for the
$v$--channel.  We also observe that the $c$--channel has sizable
contributions to both $S_z=\pm1/2$ and $S_z=\pm3/2$ amplitudes.  

From these observations, our diagram calculation 
indicates that the $S_z=\pm3/2$ contribution is significant  
as shown in Ref~\cite{Barber:1980zv}.  However, most of the 
$S_z=\pm3/2$ contribution comes from the $c$--channel, not from the
$v$--channel as suggested in Ref.~\cite{Barber:1980zv}.  We also find 
that the sizable $S_z=\pm1/2$ contributions are produced from the
$c$-- and $t$--channels.  Therefore, in order to reproduce a nearly
zero value of the ratio of
$(S_z=\pm1/2)/(S_z=\pm1/2)$~\cite{Barber:1980zv}, we need a more  
suppression factor in the $t$--channel, which is the major source of 
the $S_z=\pm1/2$ contribution in the $\Lambda^*$ photoproduction.  
\section{Summary and Conclusion}
In the present work, we investigated the $\Lambda^*(1520,3/2^-)$
photoproduction via the $\gamma N\to K\Lambda^*$ reaction.  We
employed the Rarita-Schwinger formalism for describing the spin-3/2
particle for a relativistic description. Using the
effective Lagrangians for the Born diagrams, we constructed the invariant
amplitudes for the reaction. We have investigated carefully the
dependence on the model parameters 
including various meson-baryon coupling constants and form
factors. Numerical results were then presented using reasonable sets
of the parameters for relatively low energy region $E_{\gamma}\lsim 3$
GeV. Important results are then summarized in Table.~\ref{table1}
where the features of the $\gamma p\to K^+\Lambda^*$ and  $\gamma n\to
K^0\Lambda^*$  are shown.
\begin{table}[tbh]
\begin{center}
\begin{tabular}{|c|c|c|}
\hline
Reactions&$\gamma p\to K^+\Lambda^*$ &$\gamma n\to K^0\Lambda^*$\\
\hline
$\sigma$&$\sim900\,nb$&$\sim30\,nb$\\
${d\sigma}/{d(\cos\theta)}$&Forward peak& Bump at
$\sim45^{\circ}$\\
${d\sigma}/{dt}$&Good&No data\\
\hline
\end{tabular}
\caption{Summary of the results.}
\label{table1}
\end{center}
\end{table}   
Turning to the helicity dependence, though the
contribution of the $S_z=\pm3/2$ was dominant, it was not directly related
to the $K^*$--exchange dominance as indicated in
the previous experimental analysis.  We hope that the present work
will provide a 
guideline to understanding the structure and reactions for $\Lambda(1520)$.  
\section*{Acknowledgments}
We thank T.~Nakano, H.~Toki, A.~Titov , E.~Oset and M.~J.~Vicente Vacas for
fruitful discussions and 
comments. The work of S.I.N. has been supported by
the scholarship from the Ministry of Education, Culture,
Science and Technology of Japan. The works of A.H. and S.I.N. are
partially supported by  
the collaboration program of RCNP, Osaka Univ., Japan and IFIC, Valencia
Univ., Spain. The work of A.H. is also supported in part by the Grant
for Scientific Research ((C) No.16540252) from the Education, Culture,
Science and Technology of Japan. The works of H.C.K. and S.I.N. are
supported by the Korean Research Foundation (KRF--2003--070--C00015). 
\section*{Appendix}
\subsection{Rarita-Schwinger vector-spinor}
We can write the RS vector-spinors according to their spin states as follows: 
\bee
u^{\mu}(p_{2},\frac{3}{2})&=&e^{\mu}_{+}(p_{2})u(p_{2},\frac{1}{2}),\nn\\
u^{\mu}(p_{2},\frac{1}{2})&=&\sqrt{\frac{2}{3}}e^{\mu}_{0}(p_{2})u(p_{2},\frac{1}{2})
+\sqrt{\frac{1}{3}}e^{\mu}_{+}(p_{2})u(p_{2},-\frac{1}{2}),\nn\\
u^{\mu}(p_{2},-\frac{1}{2})&=&\sqrt{\frac{1}{3}}e^{\mu}_{-}(p_{2})u(p_{2},\frac{1}{2})
+\sqrt{\frac{2}{3}}e^{\mu}_{0}(p_{2})u(p_{2},-\frac{1}{2}),\nn\\
u^{\mu}(p_{2},-\frac{3}{2})&=&e^{\mu}_{-}(p_{2})u(p_{2},-\frac{1}{2}).
\eee
Here, we employ the basis four-vectors, $e^{\mu}_{\lambda}$ which are
written by 
\bee
&&e^{\mu}_{\lambda}(p_{2})=\left(\frac{\hat{e}_{\lambda}\cdot\vec{p}_2}{M_{B}},\,\,\,\,\,\hat{e}_{\lambda}
+\frac{\vec{p}_{2}(\hat{e}_{\lambda}\cdot\vec{p}_{2})}{M_{B}(p^{0}_{2}+M_{B})}\right)\,{\rm
  with}\nn\\
&&\hat{e}_{+}=-\frac{1}{\sqrt{2}}(1,i,0),\,\,\,\,\,\,\hat{e}_{0}=(0,0,1)\,\,\,\,\,\,{\rm and}\,\,\,\,\,\hat{e}_{-}
=\frac{1}{\sqrt{2}}(1,-i,0).
\eee 
\\


\begin{thebibliography}{99}
\bibitem{experiment}
T.~Nakano {\it et al.}  [LEPS Collaboration],
Phys.\ Rev.\ Lett.\  {\bf 91}, 012002 (2003); V.~V.~Barmin {\it et al.}  [DIANA Collaboration],
Phys.\ Atom.\ Nucl.\  {\bf 66}, 1715 (2003)
[Yad.\ Fiz.\  {\bf 66}, 1763 (2003)]; S.~Stepanyan {\it et al.}  [CLAS Collaboration],
Phys.\ Rev.\ Lett.\  {\bf 91}, 252001 (2003); V.~Kubarovsky {\it et al.}  [CLAS Collaboration],
Erratum-ibid.\  {\bf 92}, 049902 (2004)
[Phys.\ Rev.\ Lett.\  {\bf 92}, 032001 (2004)]; J.~Barth {\it et al.}  [SAPHIR Collaboration],
hep-ex/0307083; A.~Airapetian {\it et al.}  [HERMES Collaboration],
Phys.\ Lett.\ B {\bf 585}, 213 (2004).
\bibitem{Nakano:chiral05}
T.~Nakano, talks in the international workshop Chiral 05, RIKEN and in
the Japna-US workshop 2005, Osaka Univ..
February (2005). 
\bibitem{Boyarski:1970yc}
A.~Boyarski, R.~E.~Diebold, S.~D.~Ecklund, G.~E.~Fischer, Y.~Murata, B.~Richter and M.~Sands,
Phys.\ Lett.\ B {\bf 34}, 547 (1971).
\bibitem{Barber:1980zv}
D.~P.~Barber {\it et al.},
Z.\ Phys.\ C {\bf 7}, 17 (1980).
\bibitem{Barrow:2001ds}
S.~P.~Barrow {\it et al.}  [Clas Collaboration],
Phys.\ Rev.\ C {\bf 64}, 044601 (2001).
\bibitem{Rarita:mf}
W.~Rarita and J.~S.~Schwinger,
Phys.\ Rev.\  {\bf 60}, 61 (1941).
\bibitem{Read:ye}
B.~J.~Read,
Nucl.\ Phys.\ B {\bf 52}, 565 (1973).
\bibitem{Johnson:1960vt}
K.~Johnson and E.~C.~G.~Sudarshan,
Annals Phys.\  {\bf 13}, 126 (1961).
\bibitem{Velo:ur}
G.~Velo and D.~Zwanziger,
Phys.\ Rev.\  {\bf 188}, 2218, (1969).
\bibitem{Pascalutsa:1998pw}
V.~Pascalutsa,
Phys.\ Rev.\ D {\bf 58}, 096002 (1998).
\bibitem{Hoehler:gb}
G.~Hoehler, H.~P.~Jakob and R.~Strauss,
Nucl.\ Phys.\ B {\bf 39},  237(1972).
\bibitem{Nath:wp}
L.~M.~Nath, B.~Etemadi and J.~D.~Kimel,
Phys.\ Rev.\ D {\bf 3}, 2153,  (1971).
\bibitem{Hagen:ea}
C.~R.~Hagen,
Phys.\ Rev.\ D {\bf 4}, 2204  (1971).

\bibitem{Machleidt:1987hj}
R.~Machleidt, K.~Holinde and C.~Elster,
Phys.\ Rept.\  {\bf 149}, 1 (1987).
\bibitem{gourdin}
M. Gourdin, Nuovo Cimento 36, 129 (1965); and, 40A, 225 (1965). 
\bibitem{Eidelman:2004wy}
S.~Eidelman {\it et al.}  [Particle Data Group],
Phys.\ Lett.\ B {\bf 592}, 1 (2004).
\bibitem{Davidson:2001qs}
R.~M.~Davidson and R.~Workman,
arXiv:nucl-th/0101066.
\bibitem{Haberzettl:1998eq}
H.~Haberzettl, C.~Bennhold, T.~Mart and T.~Feuster,
Phys.\ Rev.\ C {\bf 58},  40 (1998).
\bibitem{nam1}
S.~I.~Nam, A.~Hosaka and H.~-Ch.~Kim, talk in the few-body workshop at RCNP (2004), hep-ph/0502143.
\bibitem{nam2}
S.~I.~Nam, A.~Hosaka and H.~-Ch.~Kim,
Phys.\ Lett.\ B {\bf 579}, 43 (2004).
\bibitem{nam3}
S.~I.~Nam, A.~Hosaka and H.~-Ch.~Kim,
arXiv:hep-ph/0403009.
\bibitem{Close:2004tp}
F.~E.~Close and J.~J.~Dudek,
Phys.\ Lett.\ B {\bf 586}, 75 (2004).
\end{thebibliography}
\end{document}